\documentclass[11pt,superscriptaddress, onecolumn, nofootinbib]{revtex4-2}

\usepackage{tikz}

\usepackage[utf8x]{inputenc}

\usepackage{amssymb,amsmath}
\usepackage{latexsym}
\usepackage{graphicx}
\usepackage[font=scriptsize,labelfont=bf]{caption}
\usepackage{subcaption}
\usepackage{enumerate}
\usepackage{float}

\usepackage{color}
\usepackage{graphics}
\usepackage{epsfig}
\usepackage{bm}        
\usepackage{hyperref}        

\usepackage{textcomp}
\usepackage{natbib}
\allowdisplaybreaks

\begin{document}

\newcommand{\dR}{\mathbb R}
\newcommand{\dC}{\mathbb C}
\newcommand{\dZ}{\mathbb Z}
\newcommand{\id}{\mathbb I}
\newtheorem{theorem}{Theorem}
\newcommand{\ud}{\mathrm{d}}
\newcommand{\mfn}{\mathfrak{n}}
\newcommand{\AB}{\textcolor{blue}}
\newcommand{\PM}{\textcolor{red}}

\title{Dirac procedure and the Hamiltonian formalism for cosmological perturbations in a Bianchi I universe}

\author{Alice Boldrin}
 \email{Alice.Boldrin@ncbj.gov.pl}

\author{Przemys\l aw Ma\l kiewicz}
\email{Przemyslaw.Malkiewicz@ncbj.gov.pl}

\affiliation{National Centre for Nuclear Research, Pasteura 7, 02-093 Warsaw, Poland} 

\date{\today}

\begin{abstract}
We apply the Dirac procedure for constrained systems to the Arnowitt-Deser-Misner formalism linearized around the Bianchi I universe. We discuss and employ basic concepts such as Dirac observables, Dirac brackets, gauge-fixing conditions, reduced phase space, physical Hamiltonian,  canonical isomorphism between different gauge-fixing surfaces and spacetime reconstruction. We relate this approach to the gauge-fixing procedure for non-perturbative canonical relativity. We discuss the issue of propagating a basis for the scalar-vector-tensor decomposition as, in an anisotropic universe, the wavefronts of plane waves undergo a nontrivial evolution. We show that the definition of a gravitational wave as a traceless-transverse mode of the metric perturbation needs to be revised. Moreover there exist coordinate systems in which a polarization mode of the gravitational wave is given entirely in terms of a scalar metric perturbation. We first develop the formalism for the universe with a single minimally coupled scalar field and then extend it to the multi-field case. The obtained fully canonical formalism will serve as a starting point for a complete quantization of the cosmological perturbations and the cosmological background.

\end{abstract}


\maketitle

\section{Introduction}
 The goal of the present paper is to derive and study the Hamiltonian formalism for cosmological perturbations in the Bianchi I universe by means of the Dirac procedure for constrained systems \cite{Dirac:1964:LQM}. In this method the physical Hamiltonian and the physical phase space are obtained by solving the constraints together with gauge-fixing conditions, and replacing the Poisson bracket with the Dirac bracket. The key property of the construction is the existence of the canonical isomorphism between any two gauge-fixing surfaces, which gives rise to gauge-invariant parameterizations of the system in terms of Dirac observables. Our result agrees with \cite{Uzan} in which the standard configuration space approach is used. However, we express the physical Hamiltonian in a form that best serves the purpose of  (affine or canonical) quantization of both the background and the perturbations in a consistent manner. We provide full canonical expressions for gauge-invariant variables including the canonical definition of a gravitational wave which differs from the isotropic case. We make use of the Dirac procedure to test various gauge-fixing conditions and their isotropic limits. We discuss the reconstruction of the full spacetime metric in terms of physical phase space variables and illustrate it with an example.

The Dirac procedure based on convenient gauge-fixing conditions has been long used to ``deparametrize" full canonical relativity.  The gauge-fixing conditions promote four canonical scalar variables to the role of internal space and time coordinates in terms of which the physical degrees of freedom and their dynamics are expressed. An implementation of this procedure by use of a special canonical transformation was given by Kucha\v r (see, e.g. \cite{Kuchar:1971xm,doi:10.1063/1.1666050,Kuchar:1991qf,Isham:1992ms,Hajicek:1999ti,Brown:1994py}). This procedure paves an ideal route towards quantization of gravity, and our approach to cosmological perturbation theory follows essentially along the same lines. Nevertheless, there are also differences between the non-perturbative and perturbative approach that should be noted. The most striking one is that, in non-perturbative canonical gravity, Dirac observables are constants of motion as the Hamiltonian itself is a constraint. Therefore, the dynamical variables cannot be expressed exclusively in terms of Dirac observables and an extra variable, the internal clock, is needed. It is assumed that the internal clock commutes with all the Dirac observables. The latter makes the symplectic structure of the physical phase space depend on the choice of internal clock and gives rise to the so-called multiple choice time problem \cite{kuchar1988,Hajicek:1999ti,0264-9381-32-13-135004,Malkiewicz:2015fqa,Alexander:2012tq}. On the other hand, in the perturbative approach the second-order Hamiltonian is not a constraint, and the respective Dirac observables are, in general, dynamical. The reason for this to happen is that the first-order gauge transformations keep the background ``time" fixed whereas the true dynamics of perturbations occurs in the evolving background. Hence, the multiple choice problem is confined to zeroth order. Nonetheless, if one is to quantize both the background and the perturbations, the choice of the background clock has to be made and it will affect the dynamics of perturbations \cite{Malkiewicz2020}. We postpone the discussion of these issues to our next papers. Here, we simply note that the perturbation Hamiltonian needs to be made consistent with the choice of background internal clock via fixing the suitable lapse function. We discuss the issue of gauge-fixing in perturbation theory in geometrical terms in Sec. \ref{kucharpert}.

The dynamics of cosmological perturbations in anisotropic backgrounds exhibits new and interesting physical effects such as couplings between two modes of a gravitational wave or between gravitational waves and scalar fields \cite{Uzan,Cho_1995}. Anisotropic backgrounds are also very interesting from a theoretical point of view. For instance, the definition of polarization modes is ambiguous in anisotropic spacetimes \cite{Cho_1995}. We propose to generalize the definition of a polarization mode to be based on the Fermi-Walker-propagated vectors tangent to the wavefronts. This definition happens to give the simplest form of the dynamics and is in agreement with the prescription given in \cite{Uzan}. Unfortunately, for quantization a different definition of the basis is more suitable for the reasons we will explain later. Other interesting theoretical points are the issues of gauge-fixing and the Dirac observables. In anisotropic spacetimes the admissible gauge-fixing conditions can be quite different from those known from isotropic spacetimes. Indeed, we provide an example of a valid gauge that does not exist in the isotropic limit. Also, the Dirac observables when expressed in kinematical phase space variables are more complex as they can mix scalar, vector and tensor modes. This includes the definition of the gravitational wave which no longer can be unambiguously associated with the transverse and traceless metric perturbation. Consequently, the geometrical meaning of physical variables can vary a lot when switching from one gauge-fixing surface to another. Another interesting point is the problem of spacetime reconstruction. Once a choice of gauge has been made it requires to go back to the kinematical phase space in order to determine the values of the Lagrange multipliers, $\delta N$ and $\delta N^i$. This issue has been discussed in \cite{Malkiewicz_2019} in the context of isotropic spacetimes. Herein, we provide the spacetime reconstruction for the flat-slicing gauge. Finally, for the most part we consider a minimally coupled canonical scalar field as the only matter component, but in Sec. \ref{multifield} the formalism is extended to the case of any number of minimally coupled scalar fields in an arbitrary potential.

The obtained canonical formalism will serve in the future as a starting point for consistent quantization of all the gravitational degrees of freedom. The future quantization of the anisotropic background with the perturbations thereon, will provide a framework for testing the hypothesis of the anisotropic primordial universe in which quantum gravity effects play an essential role. The latter include the singularity resolution, the spread, the interference and entanglement effects (for the full definition of the quantum cosmological dynamics up to first order, see \cite{malkiewicz2020dynamics}, in particular Eq. (17)), which may render the framework a physically rich and viable alternative, or a complement, to the inflationary paradigm models. It must be noted that presently available analogous quantum frameworks (see e.g. \cite{Peter_2008}) are much simpler as they usually assume primordial isotropy and neglect the quantum spread or entanglement. They also generically predict (slightly) blue-tilted primordial amplitude spectrum of density perturbations in contradiction with the CMB data \cite{Aghanim:2018eyx}. The driving idea behind this work is that alternative frameworks need to possess fewer primordial symmetries and include all the relevant quantum effects. Both the theoretical and observational considerations support this idea as the available CMB data is known to contain statistical anomalies and anisotropic inflationary models can be studied in an effort to explain them \cite{pereira}.
 
For earlier works on the cosmological perturbations in a Bianchi I universe, apart from the already mentioned \cite{Uzan}, see earlier discussions of linearized Einstein's equations in \cite{Cho_1995}, or in \cite{Tomita:1985me}, where some approximate solutions were derived. The solutions for the vacuum case (i.e. for the perturbed Kasner universe) were also studied in \cite{Kofman:2011tr}. Recently, an interesting canonical analysis of the theory, by another method, was also considered in \cite{Agullo2020}. It is worth noting that similarly the perturbation theory of anisotropically curved cosmologies, the Bianchi III and the Kantowski-Sachs models, is found to lead to interesting physical effects \cite{Franco:2017pxt}.

The outline of the paper is as follows. In Sec. \ref{expansion} we expand to second order the ADM Hamiltonian constraint for gravity with a scalar field around a Bianchi I universe. In Sec. \ref{decomposition} we resolve the ADM perturbation variables into the Fourier coefficients and then decompose them into scalar, vector and tensor modes. Since the new tensorial basis is time-dependent, in Sec. \ref{Fermi} we compute the extra Hamiltonian generated by the respective transformation. In this context we also discuss the ambiguity in the definition of the polarization modes for gravitational waves. We set the spatially flat gauge-fixing conditions and obtain the physical Hamiltonian in terms of the reduced variables. In Sec. \ref{dirac} we identify the reduced variables with the Dirac observables and discuss the key concept of the canonical isomorphism between gauge-fixing surfaces. We study a few sets of gauge-fixing conditions. We make use of the consistency equation to reconstruct the full spacetime in terms of the canonical variables in the flat-slicing gauge. In Sec. \ref{multifield} we extend the obtained results to the case of many scalar fields. We conclude in \ref{conclude}. For the readers convenience, some details of our derivations are included in the Appendices.

\section{Perturbative expansion of canonical formalism}\label{expansion}
We assume that the topology of spacetime is $\mathcal{M}\simeq \mathbb{T}^3\times \mathbb{R}$ and the line element\footnote{ The purpose of assuming $\mathcal{M}\simeq \mathbb{T}^3\times \mathbb{R}$ is to have a spatially compact universe and thereby avoid the ambiguity in the definition of the symplectic structure for background (homogeneous) variables. Recall that for spatially non-compact universes one is forced to introduce an ambiguous finite fiducial cell whose choice influences the definition of the symplectic structure.},
\begin{equation}\label{st}
ds^2=-N^2dt^2+q_{ij}(dx^i+N^idt)(dx^j+N^jdt),
\end{equation}
where $i,j=1,2,3$ are spatial coordinate indices. The spacetime is filled with a real scalar field $\phi$ in a potential $V(\phi)$. The phase space of this model includes the ADM variables, the three-metric $q_{ij}$ and the three-momenta $\pi^{ij}=\sqrt{q}(K^{ij}-Kq^{ij})$ in the geometry sector, where $K_{ij}$ is the extrinsic curvature tensor, and the scalar field $\phi$ and its momentum $\pi_{\phi}$ in the matter sector:
\begin{align}
\{q_{ij}(x),\pi^{kl}(x')\}=\delta_{(i}^{~(k}\delta_{j)}^{~l)}\delta^3(x-x'),~~~\{\phi(x),\pi_{\phi}(x')\}=\delta^3(x-x').
\end{align}
The dynamics is generated by the Hamiltonian which is a sum of first-class constraints,
\begin{align}\label{fullH}
{\bf H}=\int (N \mathcal{H}_{0} +N^i\mathcal{H}_{i})~\ud^3x,
\end{align}
which comprise the gravity and matter parts,
\begin{align}
\mathcal{H}_{0}=\mathcal{H}_{g,0}+\mathcal{H}_{m,0},~~\mathcal{H}_{i}=\mathcal{H}_{g,i}+\mathcal{H}_{m,i}~,
\end{align}
with
\begin{align}\begin{split}
\mathcal{H}_{g,0}=\sqrt{q}\bigg(-{}^3R+q^{-1}(\pi_i^{~j}\pi_j^{~i}-\frac{1}{2}\pi^2)\bigg),~~
\mathcal{H}_{g,i}=-2\sqrt{q}D_j\bigg(\frac{\pi_i^{~j}}{\sqrt{q}}\bigg)~,\\
\mathcal{H}_{m,0}=\sqrt{q}\bigg(\frac{1}{2}q^{-1}\pi_{\phi}^2+\frac{1}{2}q^{ij}\phi_{,i}\phi_{,j}+V(\phi)\bigg),~~
\mathcal{H}_{ m,i}=\pi_{\phi}\phi_{,i}~,
\end{split}
\end{align}
where $D_i$ is the spatial covariant derivative. The lapse $N$ and the shifts $N^i$ play the role of the Lagrange multipliers.

We expand the above canonical formalism in perturbations around a background Bianchi I spacetime, to be introduced below, for which the three-metric and the three-momentum components in a fixed coordinate system read $\bar{q}_{ij}$ and $\bar{\pi}^{ij}$, respectively. The perturbation variables are then given by the components $\delta q_{ij}=q_{ij}-\bar{q}_{ij}$ and $\delta{\pi}^{ij}={\pi}^{ij}-\bar{\pi}^{ij}$. The lapse and the shifts, which are not dynamical variables, are also perturbed, $N\mapsto N+\delta N$ and $N^i\mapsto N^i+\delta N^i$, where $N$ and $N^i$ are now understood as zero-order quantities. The total Hamiltonian expanded up to second order reads
\begin{equation}\label{hamtot}
{\bf H}=N\mathcal{H}_{0}^{(0)}+\int_{\mathbb{T}^3} \bigg(N\mathcal{H}_{0}^{(2)}+\delta N\delta\mathcal{H}_{0}+\delta N^i\delta\mathcal{H}_{i}\bigg)~\ud^3x,
\end{equation}
where $\mathcal{H}_{0}^{(0)}$, $\delta\mathcal{H}_{0}$, $\delta\mathcal{H}_{i}$ and $\mathcal{H}_{0}^{(2)}$ are zero, first, first and second order, respectively. The integration of the zeroth-order term is omitted as we assume $\int_{\mathbb{T}^3}\ud^3x=1$. Note that there are no linear terms in the perturbations as, by assumption, they all average to zero when integrated over all space. The expansion is defined around a non-tilted Bianchi I spacetime in which the scalar field is homogeneous\footnote{The flow of the scalar field is orthogonal to the three-surfaces of homogeneity.} and thus $\mathcal{H}_{i}^{(0)}=0$. Moreover, we can choose without loss of generality $N^i=0$ at zeroth order (see, e.g. \cite{Ellis:1968vb,Ryan:1975jw}). Also, note that the full Hamiltonian \eqref{fullH} is linear in the lapse and shift functions and so must be its expansion. Strictly speaking, the above Hamiltonian does not define a gauge system as the first-order constraints $\delta\mathcal{H}_{0}$ and $\delta\mathcal{H}_{i}$'s do not commute (weakly) beyond first order. However, the assumption that the dynamics is truncated at linear order makes the first-order constraints valid generators of gauge transformations.

\subsection{The Bianchi I model}

The background metric of the Bianchi Type I model reads:
\begin{align}
\ud s^2=-\ud t^2+\sum_ia_i^2(\ud x^i)^2,~~~a=(a_1a_2a_3)^{\frac{1}{3}},
\end{align}
where we assume that the coordinates $(x^1,x^2,x^3)\in [0,1)^3$. The background Hamiltonian,
\begin{align}\label{H0}
\mathcal{H}_{0}^{(0)}=a^{-3}\bigg(\frac{1}{2}\sum_i(a_i^2p^i)^2-\sum_{i> j}a_i^2p^ia_j^2p^j+\frac{1}{2}p_{\phi}^2+a^6V\bigg),
~~\{a_i^2,p^j\}=\delta_{i}^{~j},
~\{\phi,p_{\phi}\}=1,
\end{align}
generates the equations of motion,
\begin{align}\label{backeom}\begin{split}
\dot{p}^{i}=-\frac{1}{a^3a_{i}^2}\bigg((a_i^2p^i)^2-\sum_{j\neq (i)}a_{i}^2p^{i}a_j^2p^j+a^6V\bigg),
\quad
\dot{a}^2_{i}=a^{-3}\bigg(a_{i}^4p^{i}-\sum_{j\neq (i)}a_{i}^2a_j^2p^j\bigg),\end{split}
\end{align}
\begin{align}\label{pdot}
\dot{p}_{\phi}=-a^3V_{,\phi},~~\dot{\phi}=a^{-3}{p}_{\phi},
\end{align}
and confines the dynamics to the constraint surface, $\mathcal{H}_{0}^{(0)}=0$. Finding the solution to the above equations in general can be difficult, see \cite{Mohanty:2003bj,Folomeev:2007uw,Rybakov:2010uv,Fadragas:2013ina,Chaubey:2016qtx,Kohli:2017ogi} for some results on the Bianchi I dynamics.

\subsection{First-order constraints}
Putting $\bar{q}_{ij}=a_i^2\delta_{ij}$ and $\bar{\pi}^{ij}=p^i\delta^{ij}$, the canonical perturbation variables read
\begin{align}\label{defP0}
\delta q_{ij}=q_{ij}-a_i^2\delta_{ij}
,~~\delta\pi^{ij}=\pi^{ij}-p^i\delta^{ij},~~\delta\phi=\phi-\bar{\phi},~~\delta \pi_{\phi}=\pi_{\phi}-{p}_{\phi},
\end{align}
and satisfy the following Poisson brackets
\begin{align}\label{PB1}\begin{split}
\{\delta \phi(x),\delta \pi_{\phi}(x')\}=\delta^3(x-x'),~~\{\delta q_{ij}(x),\delta\pi^{kl}(x')\}=\delta_{(i}^{~k}\delta_{j)}^{~l}\delta^3(x-x').\end{split}
\end{align}
We obtain that the first-order scalar constraint is given by\footnote{Unless specified otherwise the Einstein convention is used from now on.}

\begin{eqnarray}\label{scx}
	\begin{split}
{\delta\mathcal{H}_{0}}&=
a^{-3}
\bigg(
2\overline{\pi}_{ij}-\overline{q}_{ij}\overline{\pi}^{k}_{~k}
\bigg)
\delta\pi^{ij}
-a^{-3}
\bigg[
\frac{1}{2}\overline{q}^{ij}\bigg(\overline{\pi}^{kl}\overline{\pi}_{kl}
-\frac{1}{2}(\overline{\pi}^{k}_{~k})^2\bigg)
-2\overline{\pi}^{i}_{~k}\overline{\pi}^{kj}
+\overline{\pi}^{ij}\overline{\pi}^{k}_{~k}\bigg]\delta q_{ij}
\\
&
-a^3\overline{q}^{ij}\overline{q}^{kl}(\delta q_{ik,jl}-\delta q_{ij,kl})+a^{-3}p_{\phi}\delta\pi_{\phi}
-\frac{a^{-3}}{4}p_{\phi}^2\overline{q}^{ij}\delta q_{ij}
+\frac{a^{3}}{2}V\overline{q}^{ij}\delta q_{ij} 
+a^{3}V_{,\phi}\delta\phi,
\end{split}
\end{eqnarray}
 and the first-order vector constraint reads
\begin{eqnarray}\label{vcx}
\delta\mathcal{H}^{i}=-2\bigg(\delta\pi^{ij}_{,j}+\overline{q}^{ij}\delta q_{kj,l}\overline{\pi}^{kl}
-\frac{1}{2}\overline{q}^{ij}\delta q_{kl,j}\overline{\pi}^{kl}\bigg)+\overline{q}^{ij} p_{\phi}\delta\phi_{,j}~,
\end{eqnarray}
where $\delta q^{ij}:=\delta q_{kl}\overline{q}^{ki}\overline{q}^{lj}$ and $\delta \pi_{ij}:=\delta \pi^{kl}\overline{q}_{ki}\overline{q}_{lj}$ (and hence $({q}^{ij})^{(1)}=-\delta q^{ij}$ and $({\pi}_{ij})^{(1)}=-\delta \pi_{ij}$).

\subsection{Second-order Hamiltonian}
The second-order scalar Hamiltonian is found to read

	\small{\begin{align}
			\begin{aligned}	\label{2cx}
				\mathcal{H}_{0}^{(2)}
				&=
				a^{-3}\bigg[\delta\pi_{ij}\delta\pi^{ij}
				-\frac{1}{2}(\delta\pi^{i}_{~i})^2
				+4\delta\pi^{j}_{~k}\overline{\pi}^{ki}\delta q_{ij}
				-\overline{\pi}^{k}_k(\delta\pi^{ij}\delta q_{ij})
				-\delta\pi^{i}_{~i}(\overline{\pi}^{ij}\delta q_{ij})
								\\
				&
				-(\delta\pi^{ij}\overline{\pi}_{ij})(\delta q_{ij}\overline{q}^{ij})
				+\frac{1}{2}\overline{\pi}^{k}_{~k}\delta\pi^l_{~l}(\delta q_{ij}\overline{q}^{ij})
				+\overline{\pi}^{jl}\overline{\pi}^{in}\delta q_{il}\delta q_{jn}
								\\
				&
				-\frac{1}{2}(\overline{\pi}^{ij}\delta q_{ij})^2
				-(\overline{q}^{ij}\delta q_{ij})(\delta q_{ln}\overline{\pi}^{n}_{~m}\overline{\pi}^{ml})
				+\frac{1}{2}\overline{\pi}^{k}_{~k}(\delta q_{ij}\overline{q}^{ij})(\delta q_{ij}\overline{\pi}^{ij})\bigg]
				\\
				&
				+a^{-3}\left[\frac{1}{8}(\delta q_{ij}\overline{q}^{ij})^2
				+\frac{1}{4}(\delta q_{ij}\delta q^{ij})\right]
				\left[(\overline{\pi}^{ij}\overline{\pi}_{ij})-\frac{1}{2}(\overline{\pi}^{k}_{~k})^2\right]
				\\
				&
				+\frac{1}{4}a^{3}(\delta q_{ij}\overline{q}^{ij})(\overline{q}^{ij}\overline{q}^{kl}\delta q_{ij,kl})
				+\frac{1}{4}a^{3}(\delta q^{ij}\overline{q}^{kl})(2\delta q_{ik,jl}-\delta q_{ij,kl}
				-2\delta q_{kl,ij})
				\\
				&
				+\frac{1}{16}a^{-3}(\delta q_{ij}\overline{q}^{ij})^2p_{\phi}^2
				+\frac{1}{8}a^{-3}(\delta q^{ij}\delta q_{ij})p^2_{\phi}-\frac{1}{2}a^{-3}(\delta q_{ij}\overline{q}^{ij})p_{\phi}\delta\pi_{\phi}+\frac{1}{2}a^{-3}(\delta\pi_{\phi})^2
				\\
				&
				+\frac{1}{2}a^3\overline{q}^{ij}\delta\phi_{,i}\delta\phi_{,j}+\frac{1}{2}a^3V_{,\phi\phi}(\delta\phi)^2+\frac{1}{2}a^3(\delta q_{ij}\overline{q}^{ij})V_{,\phi}\delta\phi
				+\frac{1}{8}a^3(\delta q_{ij}\overline{q}^{ij})^2V-\frac{1}{4}a^3(\delta q_{ij}\delta q^{ij})V.
			\end{aligned}
	\end{align}
}
\section{Mode decomposition}\label{decomposition}

It is useful to introduce the rescaled spatial metric tensor $\gamma_{ij}=a^{-2}\overline{q}_{ij}$. From now on we shall use $\gamma$ to define the duals of spatial tensors. In particular, the indices of the basic perturbation variables (\ref{defP0}) are raised and lowered with  $\gamma^{ij}=(\gamma^{-1})^{ij}$ and $\gamma_{ij}$, respectively.

Let us consider the Fourier transform of a perturbation variable, denoted by $\delta X$,
\begin{align}\label{Fourier}
\delta\check{X}(\underline{k})=\int \delta X(\overline{x})e^{-ik_ix^i}\ud^3x,
\end{align}
and respectively define its inverse. The components $k_i$ of a fixed spatial co-vector $\underline{k}$ determine the respective Fourier mode by introducing wavefronts into the coordinate space. The components of the dual vector $\overline{k}$ read $k^i=k_j\gamma^{ji}$. Note that $k^i$ are in general time-dependent as $\gamma^{ij}$ evolves. It can be shown that the Fourier transform of the basic perturbation variables (\ref{defP0}) yields the following expression for the Poisson brackets (\ref{PB1}),
\begin{align}\label{PB2}
	\begin{split}
\{\delta \check{\phi}(\underline{k}),\delta \check{\pi}^{\phi}(\underline{k}')\}=\delta_{{\underline{k}},-{\underline{k}}'},~~\{\delta \check{q}_{ij}(\underline{k}),\delta\check{\pi}^{lm}(\underline{k}')\}=\delta_{(i}^{~l}\delta_{j)}^{~m}\delta_{\underline{k},-\underline{k}'}.
\end{split}
\end{align}
The Kronecker delta arises due to the compact (toroidal) topology of the spatial sheet $\Sigma\simeq \mathbb{T}^3$. Note that $k_i=2\pi n_i$, where $n_i\in\mathbb{Z}$.

As the next step, given $\overline{k}$, we introduce an orthonormal spatial triad which includes the normalized vector $\hat{k}=\overline{k}/k$, where $k=\sqrt{k_i k_j \gamma^{ij}}$. Let us denote the remaining vectors of the triad by $\hat{v}$ and $\hat{w}$. The ambiguity in defining the vectors $\hat{v}$ and $\hat{w}$ for a given $\hat{k}$ is discussed in Sec. \ref{Fermi}. Note that  the components of $\hat{v}$ and $\hat{w}$ are in general time-dependent. Making use of a fixed triad $(\hat{k},\hat{v},\hat{w})$ we may decompose any spatial symmetric $2$-rank covariant tensor in the $\gamma$-orthogonal basis $(A^1,A^2,A^3,A^4,A^5,A^6)$, where

\begin{equation}\label{baseA}
	\begin{split}
		A_{ij}^1=\gamma_{ij}
	&,
		\quad
		A_{ij}^2=\hat{k}_i\hat{k}_j-\frac{1}{3}\gamma_{ij}
		,
		\\
		A_{ij}^3=
		\frac{1}{\sqrt{2}}
		\Big(
		\hat{k}_i\hat{v}_j+\hat{v}_i\hat{k}_j
		\Big)
		&,
		\quad
		A_{ij}^4=
		\frac{1}{\sqrt{2}}
		\Big(
		\hat{k}_i\hat{w}_j+\hat{w}_i\hat{k}_j
		\Big)
		,
		\\
		A_{ij}^5=
		\frac{1}{\sqrt{2}}
		\Big(
		\hat{v}_i\hat{w}_j+\hat{w}_i\hat{v}_j
		\Big) 
		&,
		\quad
		A_{ij}^6=
		\frac{1}{\sqrt{2}}
		\Big(
		\hat{v}_i\hat{v}_j-\hat{w}_i\hat{w}_j
		\Big).
	\end{split}
\end{equation}
The dual basis $(A_1,A_2,A_3,A_4,A_5,A_6)$ is fixed in such a way that $A_n^{ij}A^m_{ij}=\delta_{n}^m$. The new basis splits the perturbations into scalar ($A^1$, $A^2$), vector ($A^3$, $A^4$) and tensor ($A^5$, $A^6$) modes. Note that both $A^n_{ij}$'s and $A_n^{ij}$'s are in general time-dependent as $\gamma_{ij}$ and $(\hat{k}_i,\hat{v}_i,\hat{w}_i)$ evolve. The perturbations of the three-metric and the three-momentum may be expressed in the new tensorial basis via the linear transformations,
\begin{align}\label{nperts}
\delta {q}_{n}=\delta \check{q}_{ij}A_n^{ij},~~~\delta {\pi}^{n}=\delta \check{\pi}^{ij}A^n_{ij}.
\end{align}
It can be shown that the Poisson brackets (\ref{PB2}) now read
\begin{align}\label{PB3}
	\begin{split}
\{\delta \check{\phi}(\underline{k}),\delta \check{\pi}^{\phi}(\underline{k}')\}=\delta_{\underline{k},-\underline{k}'},
~~\{\delta {q}_{n}(\underline{k}),\delta{\pi}^{m}(\underline{k}')\}=\delta^m_n\delta_{\underline{k},-\underline{k}'}.
\end{split}
\end{align}
The time-dependent transformations (\ref{nperts}) generate an extra term in the second-order Hamiltonian which is discussed in Sec. \ref{Fermi}.

The variables $\delta {q}_5$, $\delta {q}_6$ describe the transverse-traceless metric perturbations and, as we will see in Sec. \ref{diracobs}, they are not gauge-invariant quantities. This led to difficulties in defining the gravitational wave as a transverse and traceless metric perturbation in \cite{Cho_1995}. The transverse-traceless metric perturbations are usually called tensor modes and we follow this nomenclature in this manuscript. On the other hand, as done in \cite{Uzan}, it is also possible to define tensor modes as certain linear combinations of transverse-traceless ($\delta {q}_5$, $\delta {q}_6$) and scalar trace and traceless ($\delta {q}_1$, $\delta {q}_2$) metric perturbations that are gauge-invariant. This definition of tensor modes can then be used to describe gravitational waves as in any isotropic universe. However, this may be also misleading as it suggests that a gravitational wave has the same geometric properties in all coordinate systems (i.e. in all gauges). This is, however, not true. We discuss this issue further in Sec. \ref{dirac}.

\subsection{Zeroth-order revisited}\label{0ord}
Let us define a mapping $P:\mathrm{T}\Sigma\rightarrow \mathrm{T}\Sigma$ in the space of spatial vectors, $P^i_{~j}:=\bar{\pi}^{ik}\bar{q}_{kj}=a^2\bar{\pi}^{ik}\gamma_{kj}$. We make use of the introduced frame to define the components of $P$ such as
\begin{align}
P_{nm}
=\hat{n}_iP^i_{~j}\hat{m}^j=a^2\bar{\pi}^{ij}\hat{n}_i\hat{m}_j
=a^2\sum_ip^i\hat{n}_i\hat{m}_i
,~~\hat{n},\hat{m}\in (\hat{k},\hat{v},\hat{w}).
\end{align}

One may show that $P$ is related to the shear tensor as follows
\begin{align}
P_{{n}{m}}=a^2\sigma_{nm}-2a^2\mathcal{H}\delta_{{n}{m}},
\end{align}
where $-\frac{1}{6a^2}Tr P=~\mathcal{H}$ is the mean conformal Hubble rate. The shear is defined as $\sigma_{{n}{m}}=\sigma_{ij}\hat{n}^i\hat{m}^j$, where
\begin{align}\label{shear}
\sigma_{ij}
=\frac{1}{2}\frac{d}{d\eta}\bigg(\frac{a_i^2}{a^2}\bigg)\delta_{ij}
=
\frac{a_i^2}{a^2}\bigg(\mathcal{H}_i-\mathcal{H}\bigg)\delta_{ij},
\end{align}
and $\mathcal{H}_i=a_i^{-1}\frac{d a_i}{d \eta}$ are the directional conformal Hubble rates. The Hamiltonian constraint (\ref{H0}) becomes
\begin{align}
\mathcal{H}_0^{(0)}=a^{-3}
\left((TrP^2)-\frac{1}{2}(Tr P)^2+\frac{1}{2}p_{\phi}^2+a^6V
\right).
\end{align}
The isotropic limit is obtained by taking $P_{kk}=P_{vv}=P_{ww}=\frac{Tr P}{3}$,  and $P_{kv}=P_{kw}=P_{vw}=0$, from which we obtain $(TrP^2)=\frac{(TrP)^2}{3}$.

\subsection{First-order constraints revisited}

We express first-order constraints (\ref{scx}) and (\ref{vcx}) in the new basis $(\delta {q}_{n},\delta {\pi}^{m})$ defined in Eq. (\ref{nperts}). We find the gravity scalar constraint:
{\small\begin{align}
	\begin{aligned}
		\delta\mathcal{H}_{g,0}
		&=
		-\frac{1}{3}a^{-1}(Tr P)\delta\pi^1
		+a^{-1}[3P_{kk}
		-(Tr P)]\delta\pi^2
		+a^{-1}2\sqrt{2}P_{kv}\delta\pi^3
		+a^{-1}2\sqrt{2}P_{kw}\delta\pi^4
		\\
		&\quad
		+a^{-1}2\sqrt{2}P_{vw}\delta\pi^5
		+a^{-1}\sqrt{2}(P_{vv}
		-P_{ww})\delta\pi^6
		+\frac{1}{2}a^{-5}[(Tr P^2)
		-\frac{1}{2}(Tr P)^2]\delta q_1
		\\
		&\quad
		+\frac{1}{3}a^{-5}\{-2(Tr P^2)
		+6(P_{kk}^2+P_{kv}^2+P_{kw}^2)
		-(Tr P)\left[3P_{kk}-(Tr P)\right]\}\delta q_2
		\\
		&\quad
		+\sqrt{2}a^{-5}[2P_{kw} P_{vw}+P_{kv} (P_{vv}-P_{ww})+P_{kk} P_{kv}]\delta q_3
		\\
		&\quad
		+\sqrt{2}a^{-5}[2P_{kv}P_{vw}-P_{kw}(P_{vv}-P_{ww})+P_{kk}P_{kw}]\delta q_4
		\\
		&\quad
		+\sqrt{2}a^{-5}[2P_{kv}P_{kw}-2P_{kk}P_{vw}
		+(Tr P)P_{vw}]\delta q_5
		\\
		&\quad
		+\frac{1}{\sqrt{2}}a^{-5}[2(P_{kv}^2-P_{kw}^2)
		-2P_{kk}(P_{vv}-P_{ww})
		+(Tr P)(P_{vv}-P_{ww})
		]\delta q_6
		\\
		&\quad
		-2a^{-1}k^2(\delta q_1
		-\frac{1}{3}\delta q_2),
		\end{aligned}
\end{align}}
and the gravity vector constraint:

{	\small{\begin{align}
		\begin{aligned}
		\delta\mathcal{H}^{~i}_g=&
		-2ik\hat{k}^i
		\bigg[\frac{1}{3}\delta\pi^1
		+\delta\pi^2+a^{-4}(P_{kk}
		-\frac{1}{2}(Tr P))\delta q_1
		+\frac{a^{-4}}{6}(P_{kk}
		+(Tr P))\delta q_2
		\\
		&\qquad
		-\frac{a^{-4}}{\sqrt{2}}P_{vw}\delta q_5
		-\frac{a^{-4}}{\sqrt{2}}\frac{P_{vv}-P_{ww}}{2}\delta q_6\bigg]
		\\
		&
		-2ik\hat{v}^i
		\left[\frac{1}{\sqrt{2}}\delta\pi^3
		+a^{-4}P_{kv}
		\left(
		\delta q_1
		-\frac{1}{3}\delta q_2
		\right)
		+\frac{a^{-4}}{\sqrt{2}}P_{kk}\delta q_3
		+\frac{a^{-4}}{\sqrt{2}}P_{kw}\delta q_5
		+\frac{a^{-4}}{\sqrt{2}}P_{kv}\delta q_6
		\right]
		\\
		&
		-2ik\hat{w}^i
		\left[\frac{1}{\sqrt{2}}\delta\pi^4
		+a^{-4}P_{kw}
		\left(
		\delta q_1
		-\frac{1}{3}\delta q_2
		\right)
		+\frac{a^{-4}}{\sqrt{2}}P_{kk}\delta q_4
		+\frac{a^{-4}}{\sqrt{2}}P_{kv}\delta q_5
		-\frac{a^{-4}}{\sqrt{2}}P_{kw}\delta q_6
		\right].
		\end{aligned}
	\end{align}}}

The matter scalar and vector constraints read
\begin{eqnarray}
\delta\mathcal{H}_{m,0}=a^{-3}p_{\phi}\delta\pi_{\phi}-\frac{3}{4}a^{-5}p_{\phi}^2\delta q_{1}+\frac{3}{2}aV\delta q_{1} +a^{3}V_{,\phi}\delta\phi,~~\delta\mathcal{H}_{m}^{i}=ia^{-2}k^ip_{\phi}\delta\phi.
\end{eqnarray}
We make use of the identity $\hat{k}^i \hat{k}_j+\hat{v}^i \hat{v}_j+\hat{w}^i \hat{w}_j=\delta^{i}_{~j}$ to introduce  $\delta \mathcal{H}_k=\delta\mathcal{H}^{~i}_g\hat{k}_i+\delta\mathcal{H}_{m}^{i}\hat{k}_i$, $\delta \mathcal{H}_v=\delta\mathcal{H}^{~i}_g\hat{v}_i+\delta\mathcal{H}_{m}^{i}\hat{v}_i$ and $\delta \mathcal{H}_w=\delta\mathcal{H}^{~i}_g\hat{w}_i+\delta\mathcal{H}_{m}^{i}\hat{w}_i$. Accordingly, we define $\delta N^{k}=\delta N_i\hat{k}^i$, $\delta N^{v}=\delta N_i\hat{v}^i$ and $\delta N^{w}=\delta N_i\hat{w}^i$.

\subsection{The Fermi-Walker basis}\label{Fermi}

The frame $(\hat{k},\hat{v},\hat{w})$ is ambiguous as one has the freedom to rotate the frame along the $\hat{k}$-axis. The choice of the frame determines the definition of the polarization modes for vector and tensor perturbations. In an isotropic universe, like in the Minkowski spacetime, the definition of the two polarization modes is naturally guided by the requirement that the modes must dynamically decouple from each other. However, in an anisotropic universe decoupling is, in general, not possible. Nevertheless, we may still seek the simplest dynamical law possible. 

Let us first note that the Fourier transform fixes a foliation of the spatial coordinate space $(x^1,x^2,x^3)$ with the wavefronts of plane waves. In the physical space, due to the anisotropic dynamics, the wavefronts are not fixed but are being continuously tilted and anisotropically contracted or expanded. From the physical point of view, and which as a matter of fact is a straightforward generalization of the isotropic case, the natural choice of the vectors $(\hat{v},\hat{w})$ is to assume that they are Fermi-Walker-transported along the (future-oriented) null vector field $\vec{p}$. The spatial component of $\vec{p}$  is dual to the wavefront $\underline{k}$ of the gravitational wave (we work with $\ud s^2=-\ud\eta^2+\gamma_{ij}\ud x^i\ud x^j$), that is,
\begin{align}
\vec{p}=\overline{k}+|\overline{k}|\partial_{\eta}~,
\end{align}
where $\nabla_{\vec{p}}\vec{p}=0$. Note that $\vec{p}$ may be identified with a tangent to a bundle of null geodesics that in the eikonal approximation (i.e., for large wavenumbers) are associated with rays of gravitational waves. This construction is the so-called Sachs basis \cite{sachsbasis} that, in the context of the propagation of light in the Bianchi I universe, was also considered in \cite{Fleury:2014rea}. 

A Fermi-Walker-propagated vector field $\vec{E}$ is given by $\nabla_{\vec{p}}\vec{E}=0$,
\begin{align}
\frac{\ud E^0}{\ud\lambda}=-k^i\sigma_{ij}E^j,~~\frac{\ud E^j}{\ud\lambda}=-|\overline{k}|\sigma^j_{~i}E^i,
\end{align}
where $\lambda$ is an affine parameter. Even if $\vec{E}$ belongs initially to the plane $(\hat{v},\hat{w})$, it will eventually develop longitudinal and temporal components. Therefore, we project the covariant derivative onto the plane. We assume $^{\bot}\nabla_{\vec{p}}\vec{E}=0$, where $^{\bot}$ denotes the orthogonal projection onto the plane $(\hat{v},\hat{w})$,
\begin{align}
\frac{\ud E^j}{\ud\eta}=-\sigma^j_{~i}E^i+\hat{k}^j\sigma_{k i}E^i,
\end{align}
where the right-hand side\footnote{We identify $\sigma_{ki}=\hat{k}^j\sigma_{ji}$. Analogously defined are the quantities $\sigma_{kk}$, $\sigma_{vw}$, etc.} is projected onto the plane and  the affine parameter is replaced with conformal time via the relation $|\overline{k}|\ud\lambda=\ud \eta$. Let us assume that $(\hat{v},\hat{w})$ form a pair of Fermi-Walker-propagated vectors. Then making use of the fact that $(\hat{k},\hat{v},\hat{w})$ is an orthogonal spatial basis we obtain,
\begin{align}\label{fermilaw}
\frac{\ud \hat{v}^j}{\ud\eta}=-\sigma_{{v}{v}}\hat{v}^j-\sigma_{{v}{w}}\hat{w}^j,~~\frac{\ud \hat{w}^j}{\ud\eta}=-\sigma_{{w}{w}}\hat{w}^j-\sigma_{{w}{v}}\hat{v}^j,
\end{align}
or making use of the operator $P$ and cosmic time,
\begin{align}\label{fermilaw2}\begin{split}
\frac{\ud \hat{v}^j}{\ud t}=-a^{-3}(P_{{v}{v}}-\frac{1}{3}TrP)\hat{v}^j-a^{-3}P_{{v}{w}}\hat{w}^j,~~\frac{\ud \hat{w}^j}{\ud t}=-a^{-3}(P_{{w}{w}}-\frac{1}{3}TrP)\hat{w}^j-a^{-3}P_{{w}{v}}\hat{v}^j.
\end{split}\end{align}
The time derivatives of the dual vectors $\hat{v}_j$ and $\hat{w}_j$ are now easily determined from the time derivatives of the metric components $\gamma_{ij}$.\\

The formulae (\ref{fermilaw}) or (\ref{fermilaw2}) may be used to compute the time derivatives of the tensor basis $A^n_{ij}$ of Eq. (\ref{baseA}). In particular we are able to compute the extra Hamiltonian generated by the time dependent canonical transformation \eqref{nperts} from the coordinate-based to triad-based perturbation variables. The symplectic form is transformed as follows:
\begin{align}
\ud  \check{q}_{ij}\wedge\ud \check{\pi}^{ij}=\ud  \delta {q}_{n}\wedge\ud \delta {\pi}^{n}+\ud t \wedge\ud \left(\frac{\ud A^n_{ij}}{\ud t} A_m^{ij}\delta {q}_{n}\delta {\pi}^{m}\right),
\end{align}
leading to the extra Hamiltonian $H_{ext}=-\frac{\ud A^n_{ij}}{\ud t} A_m^{ij}\delta {q}_{n}\delta {\pi}^{m}$. The full expression for $H_{ext}$ and its reduced form in the flat slicing gauge\footnote{This is the gauge that we use in next section, that is, $\delta q_1=\delta q_2=\delta q_3=\delta q_4=0$ (see Sec. \ref{kucharpert}).} are given in Appendix \ref{NRHam0}.
The full second-order Hamiltonian (\ref{2cx}) written in terms of $(\delta {q}_{n},\delta {\pi}^{n})$, $n=1,\dots,6$, is given in Appendix \ref{NRHam}. In the following computation we will not need the full expression for we are going to impose gauge-fixing conditions that will considerably simplify Eq. (\ref{2cx}).

\section{The physical Hamiltonian}\label{physical}
In deriving the physical Hamiltonian we shall follow the Dirac procedure for constrained systems. Let us briefly describe the main steps in this procedure. We first set 4 gauge-fixing conditions $\delta c_{1}=0$, $\delta c_{2}=0$, $\delta c_{3}=0$, $\delta c_{4}=0$, which together with the initial 4 first-class constraints $\delta\mathcal{H}_{0}=0$, $\delta\mathcal{H}_{k}=0$, $\delta\mathcal{H}_{v}=0$, $\delta\mathcal{H}_{w}=0$, form a set of 8 second-class constraints. Let denote them collectively by $\delta C_{\rho}=0$, where
\begin{align}
\delta C_{\rho}=\{\delta c_{1}, \delta c_{2}, \delta c_{3}, \delta c_{4}, \delta\mathcal{H}_{0}, \delta\mathcal{H}_{k}, \delta\mathcal{H}_{v}, \delta\mathcal{H}_{w}\}.
\end{align} 
As a set of second-class constraints they form an invertible matrix of the commutation relations,
\begin{align}\label{invB}
\textrm{det} \{\delta C_{\rho},\delta C_{\sigma}\}\neq 0.
\end{align}
Provided that the above condition is satisfied, we introduce the Dirac bracket,
\begin{align}\label{db1}
\{\cdot,\cdot\}_D=\{\cdot,\cdot\} -\{\cdot,\delta C_{\rho}\}\{\delta C_{\rho},\delta C_{\sigma}\}^{-1}\{\delta C_{\sigma},\cdot\}.
\end{align}
The definition of the Dirac bracket clearly depends on the choice of the gauge-fixing conditions (we elaborate on this point in the next section). Next we impose the second-class constraints on the second-order Hamiltonian strongly, which is done by removing the redundant (i.e., dependent) dynamical variables. This yields the physical Hamiltonian, 
\begin{align}\label{physred}
H_{phys}=\bigg(N\mathcal{H}_{0}^{(2)}+\delta N\delta\mathcal{H}_{0}+\delta N^k\delta\mathcal{H}_{k}+\delta N^v\delta\mathcal{H}_{v}+\delta N^w\delta\mathcal{H}_{w}\bigg)\bigg|_{\delta C_{\rho}= 0}= N\mathcal{H}_{0}^{(2)}\bigg|_{\delta C_{\rho}= 0}.
\end{align}
The Hamilton equations in the gauge-fixing surface for any basic observable $\mathcal{O}$ are generated by the physical Hamiltonian via the Dirac bracket,
\begin{align}
\dot{\mathcal{O}}=\left\{\mathcal{O},N\mathcal{H}_{0}^{(2)}\big|_{\delta C_{\rho}= 0}\right\}_D.
\end{align}
The basic property of the above dynamics is that $\delta \dot{C}_{\rho}=0$ for all $\rho$.

\subsection{Gauge-fixing conditions}\label{kucharpert}
In what follows we impose a set of gauge-fixing conditions called the ``spatially flat slicing" gauge,
\begin{align}\label{gauge}
\delta c_1:=\delta q_1,~~\delta c_2:=\delta q_2,~~\delta c_3:=\delta q_3,~~\delta c_4:=\delta q_4.
\end{align}
Considering the perturbation of the three-curvature $\delta({}^3R)$ given in Appendix \ref{geoquant}, it is easy to see that in the above gauge $\delta({}^3R)=0$ on constant-time slices. The initial first-class constraints can be used to determine the canonical momenta, conjugate to \eqref{gauge},
\small{	\begin{align}
		\begin{aligned}\label{firtclassconst}
		&\delta\pi^1=
		\frac{2\sqrt{2}P_{vw}}{P_{kk}}\delta\pi^5
		+\frac{\sqrt{2}(P_{vv}-P_{ww})}{P_{kk}}\delta\pi^6
		+\frac{\sqrt{2}a^{-4}}{P_{kk}}
		\bigg(
		\frac{1}{2}P_{vw}[(Tr P)-P_{kk}]
		-2P_{kv} P_{kw}
		\bigg)\delta q_5
		\\
		&
		\quad
		+\frac{a^{-4}}{\sqrt{2}P_{kk}}\bigg[
		\bigg(\frac{Tr P}{2}-\frac{7}{2}P_{kk}\bigg)(P_{vv}-P_{ww})-2(P_{kv}^2-P_{kw}^2)
		\bigg]\delta q_6+\frac{a}{P_{kk}}\delta\mathcal{H}_{m,0}
		\\
		&
		\quad
		+\sum_i \frac{1}{2ik_i P_{kk}}[3P_{kk}-(Tr P)]\delta\mathcal{H}_{m,i},
		\\
		&
		\delta\pi^2=
		-\frac{2\sqrt{2}P_{vw}}{3P_{kk}}\delta\pi^5
		-\frac{\sqrt{2}}{3P_{kk}}(P_{vv}-P_{ww})\delta\pi^6
		-\frac{\sqrt{2}a^{-4}}{3P_{kk}}
		\bigg[
		\left(
		\frac{Tr P}{2}
		-2P_{kk}\right)P_{vw}
		-2P_{kv} P_{kw}
		\bigg]\delta q_5
		\\
		&
		\quad
		-\frac{a^{-4}}{3\sqrt{2}P_{kk}}
		\bigg[
		\bigg(\frac{Tr P}{2}-2P_{kk}\bigg)(P_{vv}-P_{ww})
		-2(P_{kv}^2
		-P_{kw}^2)
		\bigg]\delta q_6
		-\frac{a}{3P_{kk}}\delta\mathcal{H}_{m,0}
		+\sum_i \frac{(Tr P)}{6ik_i P_{kk}}\delta\mathcal{H}_{m,i},
		\\
		&
		\delta\pi^3=-a^{-4}P_{kw}\delta q_5
		-a^{-4}P_{kv}\delta q_6,
		\\
		&
		\delta\pi^4=-a^{-4}P_{kv}\delta q_5+a^{-4}P_{kw}\delta q_6,
	\end{aligned}
	\end{align}}
where $P_{kk}\neq 0$. Therefore, the imposition of the 8 second-class constraints, $\delta C_{\rho}=0$, naturally leads in this case to the removal of 4 canonical pairs $(\delta q_1,\delta\pi^1,\delta q_2,\delta\pi^2,\delta q_3,\delta\pi^3,\delta q_4,\delta\pi^4)$ from the phase space. The remaining variables $(\delta q_5,\delta\pi^5,\delta q_6,\delta\pi^6,\delta \check{\phi},\delta \check{\pi}^{\phi})$ are considered physical. It is easy to see that they must form 3 canonical pairs with respect to the Dirac bracket  (\ref{db1}).

The above procedure resembles the Kucha\v r decomposition in non-perturbative canonical relativity \cite{doi:10.1063/1.1666050,Hajicek:1999ti}. In the Kucha\v r decomposition one chooses four scalar fields among the kinematical variables in such a way that canonically conjugate momenta linearize the four constraints. The scalars play the role of internal spacetime coordinates with respect to which the unconstrained, ``physical" degrees of freedom and their dynamics are expressed. A well-known example is given by the Brown-Kucha\v r dust model \cite{Brown:1994py}. This  decomposition is however impossible when the spacetime model has spatial symmetries, or approximate spatial symmetries (see \cite{Torre:1992rg} for a discussion of possible obstructions to the Kucha\v r decomposition). This is the case of cosmology in which no internal spatial coordinates exist. Therefore, the role of gauge-fixing conditions is in this case slightly different. Namely, instead of unambiguously setting internal coordinates, the first-order gauge-fixing conditions fix an embedding of the fixed background spacetime in the perturbed spacetime. They implicitly determine the embedding by imposing conditions on the difference between the perturbed and the background spacetime in terms of the three-metric and the three-momentum components. This is referred to as gauge-fixing of second kind \cite{DeWitt,Nakamura}. In this way the space and time coordinates for the perturbed spacetime are unambiguously provided by the space and time coordinates of the background spacetime.  The time coordinate of the background model is an internal variable made of the kinematical background variables such as the scalar field or the scale factor, whereas the spatial coordinates of the background model are external. The latter come from a natural parametrization of the homogeneous three-space. For instance, in a toroidal Bianchi I universe, the spatial coordinates are determined only up to a constant spatial shift if the condition $\int_{\mathbb{S}_i}\ud x^i=1$ holds for all $i$. 

A more general notion than the gauge-fixing of second kind is that of covariant gauge-fixing \cite{Hajicek:1999ht}, which is a coordinate-independent notion. It is the identification of spacetime points belonging to different spacetime solutions and the existence of an underlying background spacetime with no preferred coordinates. The covariant gauge-fixing is invariant with respect to the group of diffeomorphisms of the background spacetime. The arbitrary choice of particular coordinates on the background spacetime is called the gauge-fixing of first kind. Both the point-by-point identification and the background coordinates are provided by the Kucha\v r decomposition. In the case of perturbation theory, two extra restrictions hold. First, both space and time coordinates are already fixed (modulo the spatial shifts) on the background spacetime. Second, the identification of the spacetime points belonging to different spacetime solutions has to respect the condition of the smallness of the three-metric and three-momentum perturbations.

\subsection{Physical Hamiltonian}
The physical Hamiltonian (\ref{physred}) is obtained from the second-order Hamiltonian (\ref{2cx}) by: (i) expressing it in the Fermi-Walker-propagated basis, (ii) supplementing it with the extra Hamiltonian (\ref{ext}) yielded by the time-dependent basis transformation, and (iii) reducing it into the flat slicing gauge \eqref{gauge}. It is found to read:

	\begin{align}
		\begin{aligned}\label{hfin}
	H_{phys}&=
	\frac{\delta \pi_\phi ^2}{2 a^3}
	+
	a \delta \pi_5^2
	+a
	\delta\pi_6^2
    +\bigg(\frac{k^2	a}{2}+\tilde{U}_\phi\bigg) 
    \delta \phi ^2
	+\bigg(\frac{k^2}{4 a^3}
	+\tilde{U}_5\bigg)
	\delta q_5^2
	+\bigg(\frac{k^2}{4 a^3}+\tilde{U}_6\bigg) \delta q_6^2
	\\&
	+
	C_{\phi\phi}
	\delta\phi\delta\pi_\phi+C_{55}\delta q_5
	\delta \pi_5+C_{66}\delta q_6 \delta \pi_6
	+C_{5\phi}(\sqrt{2}a^2\delta \pi_5 \delta \phi+\frac{1}{\sqrt{2} a^2}
	\delta q_5\delta \pi_\phi)
	\\&
	+C_{6\phi}
	(\sqrt{2}a^2\delta\pi_6\delta \phi +\frac{1}{\sqrt{2} a^2}\delta q_6 \delta \pi_{\phi} )+
	C_{56}(\delta q_5\delta \pi_6+\delta q_6\delta \pi_5)
	+ 
	\tilde{C}_1 \delta q_5
	\delta q_6
	+\tilde{C}_2 \delta \phi\delta q_5
	\\&
	+\tilde{C}_3 \delta \phi\delta q_6,
\end{aligned}
\end{align}
where the zero-order coefficients are given in Appendix \ref{Hfin}. The formula for the full, unconstrained second-order Hamiltonian in the Fermi-Walker basis is provided in Appendix \ref{NRHam}.

\subsection{Mukhanov-Sasaki variables}\label{mukha}
By Mukhanov-Sasaki variables we mean gauge-invariant perturbation variables that in the limit of a flat spacetime satisfy the equation of motion for the harmonic oscillator. After a suitable redefinition of the physical variables in \eqref{hfin} we obtain a set of new variables satisfying the aforementioned condition. In particular, the kinetic terms of the physical Hamiltonian (\ref{hfin}) need to be rescaled in order to describe the Hamiltonian of a harmonic oscillator. However this would not decouple the perturbation variables from their conjugate momenta. Therefore, it is useful to perform another linear canonical transformation exclusively on the momenta that removes all the couplings between momenta and positions.

Let us start by making the rescalings:
\begin{align}\label{res}\begin{split}
\delta q_{5,6}\rightarrow \delta \tilde{q}_{5,6}=\frac{1}{\sqrt{2}a}\delta q_{5,6}&,~~~\delta \pi_{5,6}\rightarrow \delta \tilde{\pi}_{5,6}=\sqrt{2}a\delta \pi_{5,6},\\
~\delta\phi\rightarrow\delta\tilde{\phi}=a\delta\phi&,
~~~\delta\pi_{\phi}\rightarrow\delta\tilde{\pi}_{\phi}=a^{-1}\delta\pi_{\phi},
\end{split}
\end{align}
which produces an extra term in the Hamiltonian, $H_{ext}=\frac{Tr P}{6a^3}( \delta \tilde{q}_{5}\delta \tilde{\pi}_{5}+\delta \tilde{q}_{6}\delta \tilde{\pi}_{6}-\delta\tilde{\phi}\delta\tilde{\pi})$.

The canonical transformation that removes the momentum-position couplings follows from the general prescription. Given the Hamiltonian,
\begin{align}
H=\frac{1}{a}\bigg(\sum_i \frac{1}{2}p_i^2+\sum_{i,j}C_{ij}p_iq_j+\dots\bigg),~~~C_{ij}=C_{ji},
\end{align}
the transformation, $p_i\rightarrow\tilde{p}_i=p_i+C_{ij}q_j$, leads to a new form,
\begin{align}\label{Hshift}
H=\frac{1}{a}\bigg(\sum_i \frac{1}{2}\tilde{p}_i^2-\frac{a}{2}\sum_{i,j}\dot{C}_{ij}q_iq_j+\dots\bigg),
\end{align}
where $~\dot{}~$ denotes the derivative with respect to cosmological time. The second term in \eqref{Hshift} arises from the time dependence of the transformation. We apply this transformation to the Hamiltonian (\ref{hfin}) for which the coefficients $C_{ij}$ are given in Eq. (\ref{HfinCoeff}). We also need to use the equations of motion for the background variables (\ref{backeom}), or for the components of $P$ in the new basis, which are given in Appendix \ref{Appbackeom}.

We arrive at the final Hamiltonian expressed in terms of the anisotropic Mukhanov-Sasaki variables:
\begin{align}\label{HBI}\begin{split}
H_{BI}=&
\frac{N}{2a}\bigg[\delta\tilde{\pi}_{\phi}^2+\delta\tilde{\pi}_5^2+\delta\tilde{\pi}_6^2+(k^2+U_{\phi})\delta\tilde{\phi}^2+(k^2+U_5)\delta \tilde{q}_5^2+(k^2+U_6)\delta\tilde{q}_6^2\\
&+C_{1}\delta \tilde{q}_5\delta \tilde{q}_6+C_{2}\delta \tilde{q}_5\delta\tilde{\phi}+C_{3}\delta\tilde{q}_6\delta\tilde{\phi}\bigg],
\end{split}
\end{align}
where the coefficients are given in Appendix \ref{MSAppgen}.  First we notice that the tensor modes are coupled to each other with the coupling $C_1$. These modes decouple when the shear of the wavefront vanishes. Analogously the scalar fields are coupled to the tensor modes with the couplings $C_2$ and $C_3$. In this case the decoupling can happen with the vanishing of the shear of the wavefront and of one of the planes perpendicular to it, i.e., given by the normal $\hat{k}\times\hat{v}$ or $\hat{k}\times\hat{w}$. The isotropic limit can be obtained as described at the end of Sec. \ref{0ord}. As expected, in this limit $C_1=C_2=C_3=0$ and the two polarization modes of the gravitational wave and the scalar mode all decouple from each other. 

The coefficients in \eqref{HBI} are combinations of the background phase space variables and the vectors $\hat{v}$ and $\hat{w}$. However, for the purpose of quantization one needs to express the vectors $\hat{v}$ and $\hat{w}$ in terms of the background phase space variables. To this end one should solve the dynamical equations \eqref{fermilaw} (or, equivalently Eqs \eqref{fermilaw2}). This however might be very difficult. In such cases one may fix $\hat{v}$ and $\hat{w}$ explicitly. In general, this procedure gives vectors that are not Fermi-Walker-propagated and therefore, it yields a correction to the extra Hamiltonian \eqref{ext}. For instance, setting:
\begin{align}
\hat{v}^i=\frac{v^i}{\sqrt{\gamma_{lj}v^lv^j}},~~\hat{w}^i=\epsilon^i_{~jk}\hat{v}^j\hat{k}^k,
\end{align}
where $v^i$ is a constant such that $v^ik_i=0$ and $\epsilon_{ijk}$ is a totally antisymmetric tensor, we find the correction to read:
\begin{align}
H_{ext}'=2a^{-2}P_{vw}(\delta \tilde{q}_5\delta\tilde{\pi}^6-\delta \tilde{q}_6\delta\tilde{\pi}^5).
\end{align}
Now the vectors $\hat{v}$ and $\hat{w}$ are given explicitly in terms of the phase space variables and the Hamiltonian \eqref{HBI} supplemented with the above correction can be quantized.

\section{Dirac observables and the gauge-fixing equivalence}\label{dirac}

The Dirac observables are defined to be first-order kinematical phase space observables, denoted by $\delta D_i$, which weakly commute with the first-class constraints,
\begin{align}\label{defdir}
\forall_{\delta{\xi}^{\rho}}~\{\delta D_i~,\int\delta{\xi}^{\rho}\delta\mathcal{H}_{\rho}\}\approx 0.
\end{align} 
The above equality is weak and thus all the Dirac observables that coincide on the constraint surface belong to the same equivalence class. We are free to choose any representative of each equivalence class to form a space of independent gauge-invariant quantities, $\mathcal{D}$. It can be shown that (a) they form a closed algebra in the constraint surface (weakly)\footnote{To show this, one needs to use the Jacobi identity. Note that the identity and zero belong to that algebra too.}, i.e. given $\delta D_i$ and $\delta D_j$, $\{\delta D_i~,\delta D_j\}\approx \delta D_k$ for some $\delta D_k$ and (b) that this algebra can be computed in {\it any} Dirac bracket (i.e., based on {\it any} choice of gauge-fixing conditions) as well as in the Poisson bracket, i.e. 
\begin{align}\label{Dalgebra}
\{\delta D_i~,\delta D_j\}_D\approx\{\delta D_i~,\delta D_j\}.
\end{align}
It is clear that the above result does not depend on the particular choice of the representatives as they all must differ by a constraint and
\begin{align}
\{\delta D_i+{\alpha}^{\rho}\delta\mathcal{H}_{\rho}~,\delta D_j\}\approx\{\delta D_i~,\delta D_j\},
\end{align}
for any ${\alpha}^{\rho}$.

The number of (independent) Dirac observables must be equal to the number of the reduced variables parametrizing any gauge-fixing surface. Furthermore, any physical variable in the gauge-fixing surface must be equal to a Dirac observable modulo a combination of constraints and gauge-fixing conditions. Specifically, for any physical variable $\delta v_i\in\{\delta\tilde{q}_{5},\delta\tilde{\pi}_{5},\delta\tilde{q}_{6},\delta\tilde{\pi}_{6},\delta\tilde{\phi},\delta\tilde{\pi}_{\phi}\}$ there must exist such $\delta D_i$ that
\begin{align}
\delta v_i=\delta D_i+{\alpha}_i^{\rho}\delta\mathcal{H}_{\rho}+{\lambda}_i^{\mu}\delta c_{\mu},
\end{align}
for some ${\alpha}_i^{\rho}$ and ${\lambda}_i^{\mu}$. Since the Dirac bracket (\ref{db1}) is blind to the constraints and to the gauge-fixing conditions we must have
\begin{align}
\{\delta v_i~,\delta v_j\}_D\approx\{\delta D_i~,\delta D_j\}_D\bigg|_{\delta c_{\rho}=0}.
\end{align}
We conclude that there exists a canonical isomorphism between the physical variables in any gauge-fixing surface and the Dirac observables,
\begin{align}\label{iso}
\mathcal{D}\ni\delta D_i\mapsto \delta v_i\approx \delta D_i\big|_{\delta c_{\mu}= 0}~,
\end{align}
 Since this mapping is a canonical isomorphism it can be used to pull-back the physical Hamiltonian (\ref{HBI}) to the space of Dirac observables, $\mathcal{D}$. Thus there exists a unique phase space, parametrized by the Dirac observables, and with the dynamics generated by a unique Hamiltonian that is a function of the Dirac observables. The choice of gauge-fixing conditions merely gives a physical meaning to the Dirac observables in terms of the kinematical phase space variables. Finally, a suitable combination of any two such canonical isomorphisms yields a canonical isomorphism between two respective gauge-fixing surfaces,
\begin{align}
\{\delta D_i\big|_{\delta c_{\rho}=0}~,\delta D_j\big|_{\delta c_{\rho}=0}\}_D\leftrightarrow\{\delta D_i\big|_{\delta c'_{\rho}=0}~,\delta D_j\big|_{\delta c'_{\rho}=0}\}_{D'}.
\end{align}

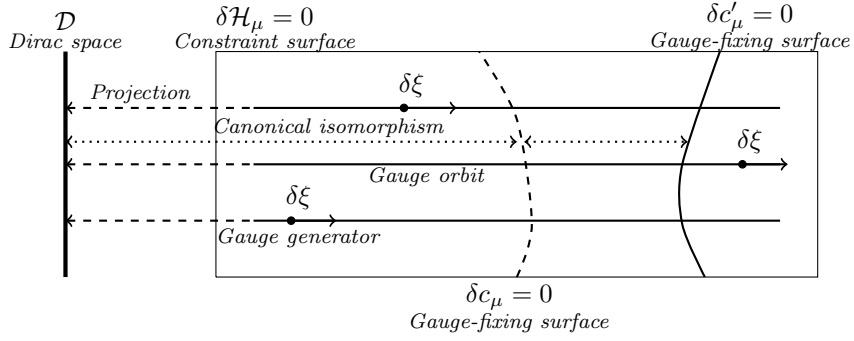
\begin{figure}

\begin{center}
\begin{tikzpicture}
\draw [color=black, ultra thick] (0,0)--(0,3)
node[label={[xshift=0cm, yshift=0cm]$\mathcal{D}$}]{}
node[label={[xshift=0cm, yshift=-0.3cm]\scriptsize{\textit{Dirac space}}}]{}
;

\draw [color=black] (2,0)--(10,0)--(10,3)--(2,3)
node[label={[xshift=0.65cm, yshift=0cm]$\delta\mathcal{H}_\mu=0$}]{}
node[label={[xshift=0.65cm, yshift=-0.25cm]\scriptsize{\textit{Constraint surface}}}]{}
--(2,0);

\draw [color=black, thick] (2.5,0.75)--(9.5,0.75);
\draw [color=black, thick] (2.5,1.5)--(9.5,1.5)
node[label={[xshift=-4.7cm, yshift=-0.55cm]\scriptsize{\textit{Gauge orbit}}}]{};
\draw [color=black, thick] (2.5,2.25)--(9.5,2.25);

\draw[color=black, thick, dashed, <-] (0,0.75)--(2.5,0.75);
\draw[color=black, thick, dashed, <-] (0,1.5)--(2.5,1.5);
\draw[color=black, thick, dashed, <-] (0,2.25)
node[label={[xshift=1cm, yshift=-0.2cm]\scriptsize{\textit{Projection}}}]{}
--(2.5,2.25);

\draw[color=black, dashed,thick] plot [smooth, tension=0.5] coordinates {(6,0) (6.2,0.7)  (6.0,2.1) (5.5,3)}
node[label={[xshift=0.4cm, yshift=-3.7cm]$\delta c_\mu=0$}]{}
node[label={[xshift=0.4cm, yshift=-3.99cm]\scriptsize{\textit{Gauge-fixing surface}}}]{};
\draw[color=black,thick] plot [smooth, tension=0.5] coordinates {(8.5,0) (8.2,0.7) (8.2,1.5)  (8.7,3)}
node[label={[xshift=0.4cm, yshift=0cm]$\delta c'_\mu=0$}]{}
node[label={[xshift=0.4cm, yshift=-0.25cm]\scriptsize{\textit{Gauge-fixing surface}}}]{};

\draw[color=black, dotted, thick,<->] (0,1.8)--(6.0,1.8)
node[label={[xshift=-2.5cm, yshift=-0.2cm]\scriptsize{\textit{Canonical isomorphism}}}]{};;
\draw[color=black, dotted, thick,<->] (6.1,1.8)--(8.27,1.8);

\filldraw[color=black] (3,0.75) circle (0.5mm)
node[label={[xshift=0.1cm, yshift=-0.1cm]$\delta \xi$}]{}
node[label={[xshift=0.1cm, yshift=-0.6cm]\scriptsize{\textit{Gauge generator}}}]{}
;
\draw[color=black, thick,->] (3,0.75)--(3.6,0.75);
\filldraw[color=black] (9,1.5) circle (0.5mm)
node[label={[xshift=0.1cm, yshift=-0.1cm]$\delta \xi$}]{};
\draw[color=black, thick,->] (9,1.5)--(9.6,1.5);
\filldraw[color=black] (4.5,2.25) circle (0.5mm)
node[label={[xshift=0.1cm, yshift=-0.1cm]$\delta \xi$}]{};
\draw[color=black, thick,->] (4.5,2.25)--(5.2,2.25);
\end{tikzpicture}
\end{center}
\caption{An illustration of the basic elements in the Dirac procedure: the constraint surface, the gauge-fixing surface, the gauge orbit, the Dirac space and the canonical isomorphism between different gauge-fixing surfaces.}\label{depict}
\end{figure}

Note that if we choose the representatives of the Dirac observables in such a way that they commute with a given set of gauge-fixing conditions,
\begin{align}\label{spD}
\{\delta D_i,\delta c_{\mu}\}=0,
\end{align} 
for all $\mu$'s, then the Dirac bracket (\ref{db1}) can be equivalently expressed as
\begin{align}
\{\cdot,\cdot\}_D=\{\cdot,\delta D_i\}\{\delta D_i,\delta D_j\}^{-1}\{\delta D_j,\cdot\}.
\end{align}
The above formula shows how the Dirac bracket operates on the kinematical phase space variables. Any variable put inside a Dirac bracket is first unambiguously associated with a Dirac observable (\ref{spD}) which coincides with that variable in a given gauge-fixing surface. Next, the resulting observable is computed in accordance with the commutation rule of Eq. (\ref{Dalgebra}) and yields a Dirac observable that is again given in the representation (\ref{spD}). The basic elements of the Dirac procedure are depicted in Fig. \ref{depict}.

\subsection{Dirac observables}\label{diracobs}
The complete set of solutions to Eq. (\ref{defdir}) in the kinematical phase space is easily found. We denote them by ${\delta D_i}$, where $i=1,\dots,10$. The explicit formulae are given in Appendix \ref{Dext}. There are 10 independent solutions in the kinematical phase space. Hence there must be six independent Dirac observables in the constraint surface. To choose them conveniently it is useful to impose the flat slicing gauge-fixing conditions $\delta q_1=\delta q_2=\delta q_3=\delta q_4=0$, which largely simplifies the expressions for ${\delta D_i}$,
\begin{align}
	\begin{split}
		&{\delta D_1}
		\big|_{GF}
		=
		{\delta q _5},
		\quad
		{\delta D_2}
		\big|_{GF}
		=	
		{\delta q _6},
		\quad
		{\delta D_3}
		\big|_{GF}
		=
		{	\delta \pi_1},
		\quad
		{\delta D_4}
		\big|_{GF}
		=
		{	\delta \pi_2},
		\quad
		{\delta D_5}
		\big|_{GF}
		=	
		{\delta \pi_3},	
	\\
		&{\delta D_6}
		\big|_{GF}
		=	
		{\delta \pi_4},
		\quad
		{\delta D_7}
		\big|_{GF}
		=
		{\delta \pi_5},
		\quad
		{\delta D_8}
		\big|_{GF}
		=
		{\delta \pi_6},
		\quad
		{\delta D_9}
		\big|_{GF}
		=
		{\delta \phi},
		\quad
		{\delta D_{10}}
		\big|_{GF}
		=	
		{\delta \pi_\phi},
	\end{split}
\end{align}
where the label $\big|_{GF}$ means ``in the gauge-fixing surface". It follows from Eq. (\ref{firtclassconst}) that the observables ${\delta D_{3}}\big|_{GF}$, ${\delta D_{4}}\big|_{GF}$, ${\delta D_{5}}\big|_{GF}$, ${\delta D_{6}}\big|_{GF}$ are not independent and may be expressed in terms of the remaining solutions. Hence we discard them. On the other hand, ${\delta D_{1}}\big|_{GF}$, ${\delta D_{2}}\big|_{GF}$, ${\delta D_{7}}\big|_{GF}$, ${\delta D_{8}}\big|_{GF}$, ${\delta D_{9}}\big|_{GF}$ and ${\delta D_{10}}\big|_{GF}$ can be easily related to the tilded variables used in the final Hamiltonian (\ref{HBI}). Since any combination of Dirac observables is a Dirac observable, we introduce a new basis in the space of Dirac observables,
\begin{align}\begin{split}
		\delta{Q}_1&=
		\frac{1}{\sqrt{2}a}{\delta D_{1}},
		~~\delta{P}_1=
		\sqrt{2}a{\delta D_{7}}
		+\frac{C_{55}+\frac{Tr P}{6a^3}}{\sqrt{2}}{\delta D_{1}}
		+\frac{C_{56}}{\sqrt{2}}{\delta D_{2}}
		+C_{5\phi}a^2 {\delta D_{9}},\\
		\delta{Q}_2&=
		\frac{1}{\sqrt{2}a}{\delta D_{2}}
		,~~\delta{P}_2=
		\sqrt{2}a{\delta D_{8}}
		+\frac{C_{56}}{\sqrt{2}}{\delta D_{1}}
		+\frac{C_{66}+\frac{Tr P}{6a^3}}{\sqrt{2}}{\delta D_{2}}
		+C_{6\phi}a^2{\delta D_{9}},\\
		\delta{Q}_{3}&=
		a{\delta D_{9}},
		~~~~~~\delta{P}_3=
		a^{-1}{\delta D_{10}}
		+\frac{C_{\phi5}}{\sqrt{2}}{\delta D_{1}}
		+\frac{C_{\phi6}}{\sqrt{2}}{\delta D_{2}}
		+(C_{\phi\phi}-\frac{Tr P}{6a^3})a^2{\delta D_{9}},
	\end{split}
\end{align}
where we used the coefficients of the Hamiltonian (\ref{hfin}). It can be verified that in the spatially flat slicing gauge the following identifications hold:
\begin{align*}
 \delta{Q}_1\big|_{GF}=\delta\tilde{q}_5,\quad \delta{Q}_2\big|_{GF}=\delta\tilde{q}_6,\quad \delta{Q}_3\big|_{GF}=\delta\tilde{\phi},\quad \delta{P}_1\big|_{GF}=\delta\tilde{\pi}_5,\quad \delta{P}_2\big|_{GF}=\delta\tilde{\pi}_6,  \quad\delta{P}_3\big|_{GF}=\delta\tilde{\pi}_{\phi}.
 \end{align*}
Making use of the respective Dirac bracket we find the canonical commutation relations,
 \[\{\delta{Q}_i(\underline{k}),\delta{P}_i(\underline{l})\}_D=\delta_{\underline{k},-\underline{l}}.\] 
 Finally, we pull-back the Hamiltonian (\ref{HBI}) to the space of Dirac observables with the mapping (\ref{iso}) and obtain
\begin{align}\label{BIn}\begin{split}
H_{BI}=&
\frac{N}{2a}\bigg[
\delta{P}_{1}^2+\delta{P}_2^2+\delta{P}_3^2
+(k^2+U_{\phi})\delta{Q}_3^2+(k^2+U_5)\delta{Q}_1^2+(k^2+U_6)\delta{Q}_2^2
\\&
+C_{1}\delta{Q}_1\delta{Q}_2+C_{2}\delta{Q}_1\delta{Q}_3+C_{3}\delta{Q}_2\delta{Q}_3\bigg]. \end{split}
\end{align}
The coefficients are given in Appendix \ref{MSAppgen}. The above Hamiltonian generates dynamical equations for all Fourier modes. Any solution is uniquely determined by the specification of the positions (describing the three-surface) and the momenta (describing the extrinsic-curvature) at initial moment of time. Note that there exists another way in which only the positions, i.e, the three-surfaces, are specified at any two fixed moments of time. It follows from the fact that the respective reduced Lagrangian (and the respective action) can be obtained in a straightforward manner from the Hamiltonian (\ref{BIn}). Hence any solution can be determined via fixing the boundary conditions and applying the least action principle. This is in complete agreement with the full GR case \cite{Baierlein:1962zz}. Obviously, the choice of gauge at the boundaries is necessary for the complete specification of the final and initial three-surfaces. The choice of the gauge at the intermediate three-surfaces is arbitrary and so is the coordinate system of the intermediate spacetime.

The basic Dirac observables expressed in terms of the kinematical phase space variables read
{\footnotesize\begin{align}\label{inddir}
	\begin{aligned}
\delta Q_1&=
\bm{\frac{1}{\sqrt{2} a}\delta q_5}
+{\frac{2P_{vw}}{a P_{kk}}}
{(\delta q_1-\frac{1}{3}\delta q_2)},
\\
 \delta Q_2&=
 \bm{\frac{1}{\sqrt{2} a}\delta q_6}
 +{\frac{P_{vv}-P_{ww}}{a P_{kk}}}
 {(\delta q_1-\frac{1}{3}\delta q_2)},
 \\
  \delta Q_3&=
  \bm{a\delta\phi+\frac{p_{\phi}}{aP_{kk}}(\delta q_1-\frac{1}{3}\delta q_2)},
  \\
\delta P_1&=
\bm{\sqrt{2} a\delta\pi_5+\frac{\frac{5}{6} (Tr P)- P_{kk}}{\sqrt{2} a^3}\delta q_5}
-{\frac{2P_{vw}}{\sqrt{2} a^3 P_{kk}}}
e{\left(\frac{P_{vv}-P_{ww}}{2}\delta q_6+P_{vw}\delta q_5\right)},
\\
&+{\mathcal{F}(P_{vw},P_{kv} P_{kw})}
{(\delta q_1-\frac{1}{3}\delta q_2)}
-{\frac{P_{vw}}{a^3P_{kk}}}
{\left(3P_{kk}\delta q_1 +a^2 p_{\phi}\delta\phi\right)}
+\frac{\sqrt{2}}{a^3}\left(P_{kw} \delta q_3+P_{kv}\delta q_4\right)
\\
 \delta P_2&=
 \bm{\sqrt{2} a\delta\pi_6+\frac{\frac{5}{6} (Tr P)- P_{kk}}{\sqrt{2} a^3}\delta q_6}
  -\frac{P_{vv}-P_{ww}}{\sqrt{2} a^3 P_{kk}}\left(\frac{P_{vv}-P_{ww}}{2}\delta q_6+P_{vw}\delta q_5\right)
 \\
 &+{\mathcal{F}\left(\frac{P_{vv} - P_{ww}}{2},\frac{P_{kv}^2-P_{kw}^2}{2}\right)}(\delta q_1-\frac{1}{3}\delta q_2)
 -{\frac{P_{vv}-P_{ww}}{2a^3P_{kk}}}\left(3P_{kk}\delta q_1 +a^2 p_{\phi}\delta\phi\right)
 +\frac{\sqrt{2}}{a^3}\left(P_{kv} \delta q_3-P_{kw}\delta q_4\right)
 \\
 \delta P_3&=
 \bm{\frac{1}{a}\delta\pi_{\phi}-\frac{(Tr P) P_{kk}+3 p_{\phi}^2}{6 a P_{kk}}\delta\phi-\frac{3p_{\phi}}{2 a^3}\delta q_1+\frac{2 (Tr P) P_{kk} p_{\phi}-6 a^6 P_{kk} V_{,\phi}-3 p_{\phi}^3}{6 a^3 P_{kk}^2}(\delta q_1-\frac{1}{3}\delta q_2)}
 \\
 &-\frac{ p_{\phi}}{\sqrt{2} a^3 P_{kk}}\left(\frac{P_{vv}-P_{ww}}{2}\delta q_6+P_{vw}\delta q_5\right)
 -\frac{p_{\phi} \left((P_{vv}-P_{ww})^2+4 P_{vw}^2\right)}{2 a^3 P_{kk}^2}(\delta q_1-\frac{1}{3}\delta q_2),
\end{aligned}
\end{align}}
where $\mathcal{F}(X,Y)=\frac{4}{a^3 P_{kk}}Y-\frac{4 \left(P_{vw}^2+ (\frac{P_{vv} - P_{ww}}{2})^2\right) + p_{\phi}^2-2P_{kk}(P_{kk}+\frac{Tr P}{3})}{a^3 P_{kk}^2}X$. The bold terms are present also in isotropic spacetimes, whereas the remaining ones are peculiar to anisotropic spacetimes. In Appendix \ref{geodirac} we give the geometric meaning of the Dirac observables, which can be used when imposing gauge-fixing conditions on geometrical quantities. 

Let us make a few observations:\\
{\bf I.} The observables  $\delta Q_1$, $\delta Q_2$ , $\delta P_1$, $\delta P_2$ transform as tensors\footnote{Note that $R_{\hat{k}}({\theta})\delta Q_1=\cos(2\theta)\delta Q_1-\sin(2\theta)\delta Q_2$, $R_{\hat{k}}({\theta})\delta P_1=\cos(2\theta)\delta P_1-\sin(2\theta)\delta P_2$, $R_{\hat{k}}({\theta})\delta Q_2=\cos(2\theta)\delta Q_2+\sin(2\theta)\delta Q_1$, $R_{\hat{k}}({\theta})\delta P_2=\cos(2\theta)\delta P_2+\sin(2\theta)\delta P_1$, where $R_{\hat{k}}({\theta})$ is the rotation around $\hat{k}=\hat{v}\times\hat{w}$ by the angle $\theta$.} and  $\delta Q_3$, $\delta P_3$ as scalars under the rotations of the $\hat{v}$-$\hat{w}$ plane. Thus, the former are the tensor modes while the latter are the scalar modes in the space of Dirac observables.\\
{\bf II.}  In the isotropic limit only the terms in bold type are non-vanishing. In this limit the number of  background variables is reduced: $P_{vv} - P_{ww}=0$, $P_{vw}=0$, $P_{kv}=0$, $P_{kw}=0$, $3P_{kk}=Tr P$. The tensor perturbations of the metric and its momentum become gauge-invariant and thus they unambiguously describe the gravitational waves. The scalar Dirac observables, $\delta Q_3$ and $\delta P_3$, in the isotropic limit are exclusively made of scalar perturbations of the field, its momentum and of the metric.\\
{\bf III.}  In the anisotropic case the tensor modes $\delta Q_1$, $\delta Q_2$ , $\delta P_1$, $\delta P_2$ are made of traceless-transverse perturbations of the metric and its momentum as well as vector and scalar perturbations of the metric. Thus,  the traceless-transverse perturbations of the metric and the momentum alone are no longer gauge-invariant. The tensorial character of the ``new" terms arises due to the zeroth-order coefficients that themselves can transform as scalars, vectors or tensors. The vector perturbations are combined with the vector zeroth-order coefficients in such a way as to yield tensors. The scalar perturbations are simply multiplied by tensor zeroth-order coefficients. Note also that tensor perturbations may be multiplied by tensor zeroth-order coefficients so as to yield scalars that are again multiplied by tensor zeroth-order coefficients. These various contributions to the tensor modes were not empasized in \cite{Uzan} where different definitions were used. In Appendix \ref{Uzantensor} we compare between the two formalisms.\\
{\bf IV.}  In the anisotropic case the scalar Dirac observable $\delta P_3$ contains tensor metric perturbations that are contracted with tensorial zeroth-order coefficients so as to yield a scalar quantity. 
\\
\indent We may generalize the above findings in the following way. In anisotropic models (that may include various matter contents) one should expect the Dirac observables to be general mixtures of scalar, vector and tensor three-metric and three-momentum perturbations as well as scalar, vector and tensor zeroth-order coefficients. It is convenient to employ the A-basis \eqref{baseA} both for the perturbations and the background quantities. Let us denote background and perturbation quantities expressed in the A-basis\footnote{E.g., $X_n=X^{ij}A_{ij}^n$ and $Y_n=\delta Y_{ij}A^{ij}_n$, where $X^{ij}$ is a background contravariant tensor and $\delta Y^{ij}$ is the perturbation of a covariant tensor. Note that it is totally irrelevant whether the considered tensors are covariant or contravariant as the metric in the $n$-label is Euclidean.} by $X_n$, $Y_n$, \dots, $n=1,\dots,6$. The quadratic scalar quantities are either the products of scalar quantities ($X_1Y_1$, $X_1Y_2$, $X_2Y_2$), or the norms of vector quantities ($X_3Y_3$, $X_4Y_4$), or the norms of tensor quantities ($X_5Y_5$, $X_6Y_6$). The quadratic vector quantities are either the products of scalar and vector quantities ($X_3Y_5$, $X_1Y_4$, $X_2Y_3$, $X_2Y_4$), or the products of vector and tensor quantities ($X_3Y_5$, $X_3Y_6$, $X_4Y_5$, $X_4Y_6$). The quadratic tensor quantities are either the products of vector quantities ($X_3Y_4$, $X_3Y_3-X_4Y_4$), or the products of scalar and tensor quantities ($X_1Y_5$, $X_1Y_6$, $X_2Y_5$, $X_2Y_6$). It is clear that the ordering of these products and whether the factors are zeroth-order, first-order or both do not matter. It is straightforward to extend these rules to cubic and higher-order products.  
\subsection{Canonical isomorphism between gauge-fixing surfaces}
Above we employed a particularly useful gauge for deriving the physical Hamiltonian. As it was shown, the choice of gauge provides a physical interpretation to the dynamical variables $(\delta{Q}_i,\delta{P}_i)$ that are Dirac observables. Below we give a few examples of other choices of gauge-fixing conditions by making use of the canonical isomorphism (\ref{iso}). Note that the condition (\ref{invB}) for the validity of any gauge can be reduced to the following:
\begin{align}
\textrm{Det}~ \Lambda\neq 0,~\textrm{where}~\Lambda_{\mu\nu}=\{\delta c_{\mu},\delta\mathcal{H}_{\nu}\}.
\end{align}
The physical and geometrical quantities will be used below for defining particular gauges. Their definition in terms of canonical variables is given in Appendix \ref{geoquant}.

\subsubsection{Uniform density gauge}

As a first example let us consider a gauge known in the isotropic limit as the {\it uniform density gauge}. It assumes the vanishing of the metric density, the energy density (defined in Appendix \ref{geoquant}) and the vector metric perturbations:
\begin{align}
\delta c_1:=\delta q_1,~~\delta c_2:=\delta \rho,~~\delta c_3:=\delta q_3,~~\delta c_4:=\delta q_4.
\end{align}
The determinant of the Poisson brackets $\Lambda_{\mu\nu}$ is non-vanishing for non-zero scalar field momentum $p_{\phi}\neq 0$,
\begin{align}
\textrm{Det}~ \Lambda = -\frac{2 i k^2 p_{\phi}^2 (Tr P)}{3 a^9},
\end{align}
proving that this is a valid gauge. The physical variables, corresponding to the Mukhanov-Sasaki variables and its conjugate, read:
\begin{align}\begin{split}
  \delta Q_3\big|_{GF}&=a\delta\phi-\frac{p_{\phi}}{3aP_{kk}}\delta q_2,\\
 \delta P_3\big|_{GF}&=-\frac{P_{vw} p_{\phi}}{\sqrt{2} a^3 P_{kk}}\delta q_5-\frac{(P_{vv}-P_{ww}) p_{\phi}}{2\sqrt{2} a^3 P_{kk}}\delta q_6-\frac{(Tr P) p_{\phi}P_{kk}+3 p_{\phi}^3+6 a^6 P_{kk}V_{,\phi}}{6 a p_{\phi}P_{kk}}\delta\phi\\
 &+\frac{-2 (Tr P) P_{kk} p_{\phi}+6 a^6 P_{kk} V_{,\phi}+3 p_{\phi} \left((P_{vv}-P_{ww})^2+4 P_{vw}^2+p_{\phi}^2\right)}{18 a^3 P_{kk}^2}\delta q_2.\end{split}\end{align}
In the anisotropic universe they contain tensor modes whereas in the isotropic universe they are combinations of the scalar field and the transverse scalar metric perturbations.

\subsubsection{Longitudinal gauge}

Our next example is another well-known gauge in the isotropic case - the \textit{longitudinal gauge}. It assumes that there is no scalar shear perturbation (defined in Appendix \ref{geoquant}) and that the shift vector vanishes. The first condition reads:
\begin{align}
	\begin{split}
\delta c_1&:=\delta\pi^2+\frac{a^{-4}}{2}\bigg(P_{kk}-\frac{Tr P}{3}\bigg)\delta q_1-\frac{2\sqrt{2}a^{-4}}{9}P_{kv}\delta q_3-\frac{2\sqrt{2}a^{-4}}{9}P_{kw}\delta q_4\\
&-\frac{4\sqrt{2}a^{-4}}{9}P_{vw}\delta q_5-\frac{2\sqrt{2}a^{-4}}{9}(P_{vv}-P_{ww})\delta q_6.
\end{split}
\end{align}
Note that in the isotropic limit this simply reads $\delta c_1=\delta\pi^2$.
We fix the second condition in agreement with the isotropic limit, i.e.,
\begin{align}
\delta c_2:=\delta q_2.
\end{align}
 When combined with $\delta c_3:=\delta q_3$ and $\delta c_4:=\delta q_4$, this set of conditions yields:
\begin{align}
\textrm{Det}~ \Lambda =-\frac{2 i k^2 \left(2 \left(6 a^4 k^2+9 P_{kk}^2+18( P_{kv}^2+P_{kw}^2)+(P_{vv}-P_{ww})^2+4 P_{vw}^2\right)-15 P_{kk} Tr P+3(Tr P)^2\right)}{9 a^5},
\end{align}
which proves it to be a valid gauge.

\subsubsection{Scalar gravity-wave gauge}

Our final example is chosen to demonstrate that anisotropic spacetimes can accommodate gauges that do not exist in isotropic spacetimes. We shall call it the {\it scalar gravity-wave gauge}. It assumes the vanishing of a scalar, a vector and a tensor metric perturbation:
\begin{align}
\delta c_1=\delta q_2,~~\delta c_2:=\delta q_3,~~\delta c_3:=\delta q_4,~~\delta c_4:=\delta q_5.
\end{align}
The determinant of the Poisson brackets $\Lambda_{\mu\nu}$,
\begin{align}
\textrm{Det}~ \Lambda = -\frac{8 i \sqrt{2} k^2 P_{vw}}{ a},
\end{align}
is non-vanishing as long as $P_{vw}\neq 0$, i.e. the $A_5$-component of the wavefront shear does not vanish\footnote{This reasoning would still hold if we chose $\delta c_4:=\delta q_6$, with the only difference that in this case $\textrm{Det}~  \Lambda =- \frac{4 i \sqrt{2} k^2(P_{vv}-P_{ww})}{a}$, i.e., the $A_6$-component of the wavefront shear is assumed not to vanish. Note however that the vanishing of both tensor metric perturbations is impossible as one would have $\textrm{Det}~  \Lambda=0$}. It is clear that this gauge must cease to exist in the isotropic limit. Note that in this gauge one of the polarization modes of the gravitational wave is entirely carried by the metric density perturbation:
\begin{align}
\delta Q_1\big|_{GF}=\frac{2P_{vw}}{a P_{kk}}\delta q_1.
\end{align}
This gauge corresponds to a coordinate system in which a given tensor metric perturbation is vanishing and the gravity wave is induced by a scalar perturbation that perturbs the wavefronts with non-vanishing shear. It demonstrates that the transverse and traceless metric perturbations cannot be unambiguously identified with gravitational waves. 

\subsection{Spacetime reconstruction}
In order to obtain a complete spacetime picture as in Eq. (\ref{st}) we need to determine the values of $\delta N$, $\delta N^k$, $\delta N^v$ and $\delta N^w$ from the consistency equations,
\begin{equation}\label{stabilityEq}
\{\delta c_{\rho},{\bf H}\} \approx 0,~~~\rho=1,2,3,4,
\end{equation}
where ${\bf H}$ is the unconstrained Hamiltonian of Eq. (\ref{hamtot}). Hence for each mode $\underline{k}$ and each $\rho$, we obtain a linear algebraic equation,
\begin{equation}
N\{\delta c_{\rho},\mathcal{H}_{0}^{(0)}+\mathcal{H}_{0}^{(2)}\}+\delta N^{\mu}\Lambda_{\rho\mu} \approx 0,
\end{equation}
where $\mathcal{H}_{0}^{(2)}$ should include the extra Hamiltonian if the perturbation variables are expressed in the $A$-basis. Given a complete set of gauge-fixing conditions, the above equation is easily solved
\begin{align}
\frac{\delta N^{\mu}}{N}\approx-(\Lambda^{-1})^{\mu\rho}\{\delta c_{\rho},\mathcal{H}_{0}^{(0)}+\mathcal{H}_{0}^{(2)}\}.
\end{align}
For the particular case of the flat slicing gauge (\ref{gauge}) the consistency equations in the Fermi-Walker-propagated basis yield, in terms of the Dirac observables,
\begin{align}\begin{split}
\frac{\delta N}{N}&=
-\frac{P_{vw} }{ a P_{kk}}\delta Q_1
-\frac{P_{vv}-P_{ww}}{2 a P_{kk}}\delta Q_2
-\frac{p_\phi  }{2a P_{kk}}\delta Q_3,\\
	\frac{\delta N^k}{N}&=
\delta Q_1
\bigg(
\frac{ P_{vw}}{2 a^2} 
-\frac{ 2 P_{kv} P_{kw} }{a^2 P_{kk}}
\bigg)
+\delta Q_2
\bigg(
\frac{ P_{kw}^2}{a^2 P_{kk}} 
-\frac{ P_{kv}^2 }{a^2 P_{kk}}
+\frac{ P_{vv}-P_{ww} }{4  a^2}
\bigg)
\\&+\delta Q_3
\bigg(
\frac{ a^4 V_{,\phi} }{2 P_{kk}}
-\frac{p_\phi (TrP)  }{2 a^2 P_{kk}}
+\frac{3  p_\phi  }{4 a^2}
\bigg)
+\frac{ P_{vw}}{P_{kk}}\delta P_1
+\frac{ P_{vv}-P_{ww}}{2 P_{kk}}\delta P_2
+\frac{ p_\phi  }{2 P_{kk}}\delta P_3
,\\
\frac{\delta N^v}{N}&=
\delta Q_1
\bigg(
\frac{ 2 P_{kv} P_{vw} }{a^2 k P_{kk}}
+\frac{ 2 P_{kw}}{a^2 k}
\bigg)
+
\delta Q_2
\bigg(
\frac{ P_{kv} (P_{vv}-P_{ww}) }{a^2 k P_{kk}}
+\frac{ 2 P_{kv} }{a^2 k}
\bigg)
+\frac{ P_{kv} p_\phi  }{a^2 k P_{kk}}\delta Q_3
,\\
\frac{\delta N^w}{N}&=
\delta Q_1
\bigg(
\frac{2 P_{kw} P_{vw}}{a^2 k P_{kk}}
+\frac{ 2 P_{kv}}{a^2 k}
\bigg)
+\delta Q_2
\bigg(
\frac{ P_{kw} (P_{vv}- P_{ww})}{a^2 k P_{kk}}
-\frac{2 P_{kw} }{a^2 k}
\bigg)
+\frac{P_{kw} p_\phi  }{a^2 k P_{kk}}\delta Q_3
.\end{split}
\end{align}
Note that the lapse function $\delta N$ and the shift vector $\delta N^k$ are scalars under rotations around $\hat{k}$. It follows from the fact that the tensor Dirac observables $\delta Q_1$, $\delta Q_2$, $\delta P_1$ and $\delta P_2$ suitably multiplied by the tensor zeroth-order coefficients (i.e. forming the products $X_5Y_5$ and $X_6Y_6$, see our discussion below \eqref{inddir}) become scalars. We also note that under those rotations the shift vectors $\delta N^v$ and $\delta N^w$ transform as vectors\footnote{I.e. $R_{\hat{k}}({\theta})\delta N^v=\cos(\theta)\delta N^v+\sin(\theta)\delta N^w$, $R_{\hat{k}}({\theta})\delta N^w=\cos(\theta)\delta N^w-\sin(\theta)\delta N^v$, where $R_{\hat{k}}({\theta})$ is the rotation around $\hat{k}=\hat{v}\times\hat{w}$ by the angle $\theta$.} (as they include quadratic and cubic terms $X_3Y_5$, $X_3Y_6$, $X_4Y_5$, $X_4Y_6$, $X_3Y_5Z_5$, $X_4Y_5Z_5$, $X_3Y_6Z_6$, $X_4Y_6Z_6$). The knowledge of $\frac{\delta N}{N}$, $\frac{\delta N^k}{N}$,  $\frac{\delta N^v}{N}$,  $\frac{\delta N^w}{N}$ allows to reconstruct the full spacetime metric tensor of Eq. (\ref{st}) as a function of Dirac observables. However, the metric components (\ref{st}) themselves are not Dirac observables as the above relations are tied to the particular choice of gauge-fixing conditions (i.e. they depend on the employed coordinate system).

An alternative way of fixing gauge conditions would be to start by setting $\frac{\delta N}{N}$, $\frac{\delta N^k}{N}$,  $\frac{\delta N^v}{N}$,  $\frac{\delta N^w}{N}$ and then to solve the equation:
\begin{align}
\{\delta c_{\rho},\mathcal{H}_{0}^{(0)}+\mathcal{H}_{0}^{(2)}\}=-\Lambda_{\rho\mu}\frac{\delta N^{\mu}}{N},
\end{align}
which is a (linear) first-order partial differential equation that in general has no unique solution. This leads to the so-called residual gauge freedom. Geometrically, the choice of  $\frac{\delta N^{\mu}}{N}$ fixes the coordinates uniquely once the initial Cauchy surface is fixed. Choosing a specific solution to the above equation is therefore equivalent to setting the initial Cauchy surface in the perturbed spacetime.

\section{Multi-field case}\label{multifield}

Particle physics models allow for the possible presence of more than one scalar field in the primordial universe. The multi-field scenario offers new and attractive features in particular to the theory of inflation \cite{2009pdpL,Bassett_2006,Schutz:2013fua}. It is quite straightforward to extend the above single-field framework to the case when the matter is made of a collection of scalar fields. The matter constraints read:
\begin{align}\begin{split}
\mathcal{H}_{m,0}=\sqrt{q}\bigg(\sum_I\frac{1}{2}q^{-1}\pi_{\phi^I}^2+\sum_I\frac{1}{2}q^{ij}\phi_{~,i}^I\phi_{,j}^I+V(\dots,\phi^I,\dots)\bigg),~~
\mathcal{H}^{~i}_m=-\sum_I\pi_{\phi^I}\phi_{,}^{I~i}~,
\end{split}
\end{align}
where the index $I$ labels the real scalar fields. The Fourier components of the linearized scalar and vector matter constraints are easily found to read:
\begin{equation}
	\begin{split}
\delta\mathcal{H}_{m,0}&=\sum_I\left(a^{-3}p_{\phi^I}\delta\pi_{\phi^I}-\frac{3}{4}a^{-5}(p_{\phi^I})^2\delta q_{1}+\frac{3}{2}aV\delta q_{1} +a^{3}V_{,\phi^I}\delta\phi^I\right),\\
\delta\mathcal{H}_{m}^{i}&=\sum_Iia^{-2}k^ip_{\phi^I}\delta\phi^I.
\end{split}
\end{equation}
We proceed by essentially repeating all the steps made in the single-field case. In particular we set the spatially flat gauge-fixing conditions (\ref{gauge}). We arrive at the following generalization of the gauge-invariant Hamiltonian (\ref{HBI}):
\begin{align}\label{HBIgen}\begin{split}
H_{BI}=&
\frac{N}{2a}\bigg[
\delta{P}_{1}^2+\delta{P}_2^2+\sum_I\delta{P}_{3I}^2
+\sum_I(k^2+U_{\phi})\delta{Q}_{3I}^2+(k^2+U_5)\delta{Q}_1^2+(k^2+U_6)\delta{Q}_2^2
\\&
+C_{1}\delta{Q}_1\delta{Q}_2+\sum_IC_{2I}\delta{Q}_1\delta{Q}_{3I}+\sum_IC_{3I}\delta{Q}_2\delta{Q}_{3I}+\sum_{I>J}C_{IJ}\delta Q_{3I}\delta Q_{3J}\bigg], \end{split}
\end{align}
where the coefficients are given in Appendix \ref{MSAppgen}. The above Hamiltonian is basically of the same type as (\ref{HBI}) with the difference that there are more matter perturbations denoted by $\delta Q_{3I}$'s and that it includes couplings between them, $C_{IJ}$'s. Note that for the spatially flat gauge the following identifications hold:
\begin{align*}
 \delta{Q}_1=\delta\tilde{q}_5,\quad \delta{Q}_2=\delta\tilde{q}_6,\quad \delta{Q}_{3I}=\delta\tilde{\phi^I},\quad \delta{P}_1=\delta\tilde{\pi}_5,\quad \delta{P}_2=\delta\tilde{\pi}_6 \quad \text{and} \quad\delta{P}_{3I}=\delta\tilde{\pi}_{\phi^I}.
 \end{align*}

The complete set of Dirac observables is given in Appendix \ref{Dext}. The physical variables are related to them as follows:
 \begin{align}
 	\begin{split}
		\delta{Q}_1&=
		\frac{1}{\sqrt{2}a}{\delta D_{1}},
		~~\delta{P}_1=
		\sqrt{2}a{\delta D_{7}}
		+\frac{C_{55}}{\sqrt{2}a}{\delta D_{1}}
		+\frac{C_{56}}{\sqrt{2}a}{\delta D_{2}}
		+C_{5\phi^J}a {\delta D_{9J}},\\
		\delta{Q}_2&=
		\frac{1}{\sqrt{2}a}{\delta D_{2}}
		,~~\delta{P}_2=
		\sqrt{2}a{\delta D_{8}}
		+\frac{C_{56}}{\sqrt{2}a}{\delta D_{1}}
		+\frac{C_{66}}{\sqrt{2}a}{\delta D_{2}}
		+C_{6\phi^J}a{\delta D_{9J}},\\
		\delta{Q}_{3I}&=
		a{\delta D_{9I}},
		~~~~~~\delta{P}_{3I}=
		a^{-1}{\delta D_{10I}}
		+\frac{C_{I 5}}{\sqrt{2}a}{\delta D_{1}}
		+\frac{C_{ I6}}{\sqrt{2}a}{\delta D_{2}}
		+C_{\phi^I\phi^J}a{\delta D_{9J}}.
	\end{split}
\end{align}

\section{Conclusions}\label{conclude}

In the present work we have derived the physical phase space and the physical Hamiltonian for anisotropic cosmological theory by means of the Dirac procedure. We introduced the Fermi-Walker-propagated basis to define the tensor and vector modes and computed the extra Hamiltonian generated by the choice of that basis. We chose convenient gauge-fixing conditions and obtained the physical Hamiltonian in terms of background and perturbation canonical variables. We showed that the obtained result is valid in any gauge if the physical variables are replaced by the respective Dirac observables. We also reconstructed the full spacetime by means of canonical variables in the flat slicing gauge. Finally, we extended the obtained result to the multi-field case which may be relevant for models of the primordial universe.

The Dirac method relies on the existence of the canonical isomorphism between different gauge-fixing surfaces and provides a useful framework for studying gauge-fixing conditions. We considered a few examples of sets of gauge-fixing conditions. Some of them turned out to be  extensions of the gauge-fixing conditions used in isotropic spacetimes while others do not exist in isotropic spacetimes. For any choice of gauge-fixing conditions the Dirac observables acquire a distinct physical interpretation. In particular, a gravitational wave could be seen as a scalar perturbation of the metric. More precisely, we showed that a gravitational wave could be completely given by a wave of the perturbation of the volume density even in the absence of matter. It is interesting to note that the existence of the scalar mode of gravitational wave is well-known in extended or modified GR theories \cite{capozziello}. It has to be added though that our ``scalar" wave, unlike the one in modified GR, preserves the tensorial transformation properties due to the anisotropy of the wavefronts that is induced by the global dynamics of the model. 

In the future we shall investigate the quantization of the obtained framework. For this purpose at the end of Sec. \ref{mukha} we derived the physical Hamiltonian in a non-Fermi-Walker basis that (unlike the Fermi-Walker one) can be formulated explicitly and completely in terms of background phase-space variables. The derived formalism can be reduced to the vacuum case, that is, the Kasner universe filled with gravitational waves by putting the background scalar field and its momentum, and the perturbations thereof, to zero. This simplified setup could be studied first. A general approach to quantization of cosmological perturbations with dynamical backgrounds was given e.g. in \cite{malkiewicz2020dynamics}. The first key issues to consider are the choice of internal clock and the subsequent quantization of the zeroth-order coefficients and their dynamics. Finding the solution to the equations of motion for perturbations in a quantized anisotropic background is going to be a hard problem, and perhaps some approximations will be needed. Nonetheless we should be able to answer many interesting questions. For example, how the quantum theory depends on the choice of gauge-fixing conditions, and in particular, on the choice of internal clock. We expect that, in general, the quantization will break the diffeomorphism invariance of the classical theory (this issue has been already studied at the background level in \cite{Malkiewicz2020}). At the same time we expect that the physical predictions for the large, classical universe given by the quantum theory (such as the power spectrum of primordial perturbations) should not depend on the choice of gauge.

\begin{acknowledgments}
The authors acknowledge the support of the National Science Centre (NCN, Poland) under the research grant 2018/30/E/ST2/00370.
We would also like to thank Jan Ostrowski and Patrick Peter for helpful discussions.
\end{acknowledgments}

\appendix

\section{Extra Hamiltonian}\label{NRHam0}

Since the canonical transformation \eqref{nperts} depends on time through the new tensorial basis \eqref{baseA}, it generates an extra term in the full Hamiltonian (\ref{hamtot}). The extra term depends on the gravitational variables only, that is, $(\delta {q}_{n},\delta {\pi}^{n})$, $n=1,\dots,6$, and reads

\begin{align}\label{fullext}
&{H_{ext}}=
-\Bigg[
\delta q_1 \bigg(\frac{  (3 P_{kk}	-(TrP))}{ a^3}\delta \pi_2
	+\frac{2 \sqrt{2}  P_{kv}}{a^3}\delta \pi_3
	+\frac{2 \sqrt{2}  P_{kw}}{a^3}\delta \pi_4
	+\frac{\sqrt{2}\left( P_{vv}- P_{ww}\right)}{ a^3}\delta \pi_6 
 	+\frac{2 \sqrt{2}P_{vw}}{a^3} \delta \pi_5
 	\bigg)
	\nonumber
	\\
\nonumber
&
\quad
+\delta q_2 \bigg(\frac{ 2(3 P_{kk}	- (TrP))}{9 a^3}\delta \pi_1
	+\frac{ (3 P_{kk}	-(TrP))}{3 a^3}\delta \pi_2
	-\frac{2 \sqrt{2}  P_{kv}}{3 a^3}\delta \pi_3
	-\frac{2 \sqrt{2}  P_{kw}}{3 a^3}\delta \pi_4
		\nonumber
	\\
	\nonumber
	&
	\qquad
	-\frac{2 \sqrt{2} P_{vw}}{3 	a^3}\delta \pi_5
	+\frac{ \sqrt{2} \left( P_{ww}	- P_{vv}\right)}{3 a^3}\delta \pi_6
	\bigg)
	\\
\nonumber
&
\quad
+\delta q_3\bigg(\frac{2 \sqrt{2} P_{kv}}{3 a^3} \delta \pi_1
	+\frac{2 \sqrt{2}  P_{kv}}{a^3}\delta \pi_2
	+\frac{ ( (TrP)	-3 P_{ww})}{3 a^3}\delta \pi_3
	+\frac{P_{vw}}{a^3}\delta \pi_4
	\bigg)
	\\
	&
\quad
+\delta q_4 \bigg(\frac{2 \sqrt{2}  P_{kw}}{3 a^3}\delta \pi_1
	+\frac{2 \sqrt{2}  P_{kw}}{a^3}\delta \pi_2
	+\frac{P_{vw}}{a^3}\delta \pi_3
	+\frac{ ( (TrP)-3 P_{vv})}{3 a^3}\delta \pi_4
    \bigg)
	\\
\nonumber
&
\quad
+\delta q_5 \bigg(
		+\frac{2 \sqrt{2} P_{vw}}{3 a^3}\delta \pi_1
	-\frac{\sqrt{2} P_{vw}}{a^3}\delta \pi_2
	+\frac{2  P_{kw}}{a^3}\delta \pi_3
	+\frac{2  P_{kv}}{a^3}\delta \pi_4
	+\frac{ ( (TrP)-3 P_{kk})}{3 a^3}\delta \pi_5
\bigg)
\\
\nonumber
&
\quad
+\delta q_6 \bigg(
	\frac{\sqrt{2} \left( P_{vv}- P_{ww}\right)}{3 a^3}\delta \pi_1
	-\frac{ \left( P_{vv}	- P_{ww}\right)}{\sqrt{2} a^3}\delta \pi_2
	+\frac{2  P_{kv}}{a^3}\delta \pi_3
	-\frac{2  P_{kw}}{a^3}\delta \pi_4
	+\frac{ ((TrP)-3 P_{kk})}{3 a^3}\delta \pi_6
	\bigg)\Bigg].
\end{align}

In the spatially flat slicing gauge, the extra Hamiltonian takes the form
\small{\begin{align}\label{ext}
		\begin{aligned}
			H_{ext}=-a^{-3}\delta q_5\left(\frac{2\sqrt{2}}{3}P_{vw}\delta\pi^1-\sqrt{2}P_{vw}\delta\pi^2+2P_{kw}\delta\pi^3+2P_{vk}\delta\pi^4+\left(\frac{Tr P}{3}-P_{kk}\right)\delta\pi^5\right)\\
			-a^{-3}\delta q_6\left(\frac{\sqrt{2}}{3}(P_{vv}-P_{ww})\delta\pi^1-\frac{P_{vv}-P_{ww}}{\sqrt{2}}\delta\pi^2+2P_{kv}\delta\pi^3-2P_{kw}\delta\pi^4+\left(\frac{Tr P}{3}-P_{kk}\right)\delta\pi^6\right).
		\end{aligned}
\end{align}}

\section{Second-order constraint}\label{NRHam}

The second order constraint (\ref{2cx}) expressed in the new tensorial basis (\ref{baseA}) and supplemented with the extra term (\ref{fullext}) generated by the corresponding time-dependent canonical transformation reads

{\small
\begin{align}
	&{\mathcal{H}_{0}^{(2)}+H_{ext}}=
	-\frac{a \delta \pi_1^2}{6}
	+\frac{3 a \delta\pi_2^2}{2}
	+a \delta \pi_3^2
	+a \delta \pi_4^2
	+a \delta \pi_5^2
	+a \delta \pi_6^2
	+\frac{\delta\pi_\phi ^2}{2 a^3}
		\nonumber
	\\
	&
		+\delta\phi ^2\bigg(
	\frac{ a^3V_{,\phi\phi}}{2}
	+\frac{k^2a}{2}
	\bigg) 
	+
	\delta q_1^2\bigg(
	\frac{3 V}{8 a}
	-\frac{k^2}{2 a^3}
	+\frac{15p_\phi^2}{16 a^7}
	+\frac{ (Tr P)^2}{16a^7}
	-\frac{(Tr P^2)}{8 a^7}\bigg) 
			\nonumber
	\\
	&
	+\delta q_2^2
	\bigg(-\frac{k^2}{18 a^3}
	-\frac{P_{kk}^2}{6 a^7}
	-\frac{2 P_{kv}^2}{3a^7}
	-\frac{2 P_{kw}^2}{3 a^7}
	+\frac{p_\phi^2}{12 a^7}
	-\frac{V}{6 a}
	+
	\frac{P_{kk} (Tr P)}{3	a^7}
	-\frac{5(Tr P)^2}{36 a^7}
	+\frac{5 (Tr P^2)}{18 a^7}\bigg)
	\nonumber
	\\
	&
	+\delta q_3^2
	\bigg(
	\frac{p_\phi^2}{8 a^7}
	-\frac{(Tr P)^2}{8a^7}
	+\frac{P_{kk} P_{vv}}{a^7}
	-\frac{V}{4 a}
	+\frac{(Tr P^2)}{4a^7}
	\bigg)
	+\delta q_4^2 
	\bigg(
	\frac{p_\phi^2}{8 a^7}
	-\frac{(Tr P)^2}{8 a^7}
	+\frac{P_{kk} P_{ww}}{a^7}
	-\frac{V}{4 a}
	+\frac{(Tr P^2)}{4 a^7}
	\bigg)
	\nonumber
	\\
	&
	+\delta q_5^2 
	\bigg(
	\frac{k^2}{4 a^3}
	+\frac{p_\phi^2}{8a^7}
	-\frac{(Tr P)^2}{8 a^7}
	+\frac{P_{vv}P_{ww}}{a^7}
	-\frac{V}{4 a}
	+\frac{(Tr P^2)}{4 a^7}
	\bigg)
	\nonumber
	\\
	&
	+\delta q_6^2
	\bigg(
	\frac{k^2}{4 a^3}
	+\frac{(P_{vv}+P_{ww})^2}{4a^7}
	-\frac{P_{vw}^2}{a^7}
	+\frac{p_\phi^2}{8 a^7}
	-\frac{(Tr P)^2}{8 a^7}
	-\frac{V}{4 a}
	+\frac{(Tr P^2)}{4 a^7}
	\bigg)
	\nonumber
	\\
	&
	+
\delta q_1\bigg[
	-\frac{\sqrt{2}	P_{kv} \delta \pi_3}{a^3}
	-\frac{\sqrt{2} P_{kw} \delta \pi_4}{a^3}
	-\frac{\sqrt{2} P_{vw} \delta \pi_5}{a^3}
	+\left(\frac{P_{ww}}{\sqrt{2} a^3}
	-\frac{P_{vv}}{\sqrt{2}	a^3}\right) \delta \pi_6
	-\frac{3 p_\phi \delta\pi_\phi }{2a^5}
	-\frac{\delta \pi_1 (Tr P)}{6 a^3}
		\nonumber
	\\
	&
	+\delta \pi_2
	\bigg(
	\frac{(Tr P)}{2 a^3}
	-\frac{3P_{kk}}{2 a^3}
	\bigg)
	+\delta q_3
	\bigg(
	\frac{\sqrt{2} P_{ww}P_{kv}}{a^7}
	-\frac{ (Tr P) P_{kv}}{\sqrt{2}	a^7}
	-\frac{\sqrt{2} P_{kw} P_{vw}}{a^7}
	\bigg)
		\nonumber
	\\
	&
	+\delta q_4
	\bigg(
	\frac{\sqrt{2} P_{vv}P_{kw}}{a^7}
	-\frac{ (Tr P) P_{kw}}{\sqrt{2}	a^7}
	-\frac{\sqrt{2} P_{kv} P_{vw}}{a^7}
	\bigg)
		\nonumber
	\\
	&
	+\delta q_5
	\bigg(
	\frac{\sqrt{2} P_{kk}	P_{vw}}{a^7}
		-\frac{	(Tr P)P_{vw}}{\sqrt{2} a^7}
	-\frac{\sqrt{2} P_{kv} P_{kw}}{a^7}
	\bigg)
		\nonumber
	\\
	&
	+\delta q_6
	\bigg(
	-\frac{P_{kv}^2}{\sqrt{2} a^7}
	+\frac{P_{kw}^2}{\sqrt{2}a^7}
	-\frac{(P_{vv}^2-P_{ww}^2)}{\sqrt{2} a^7}
	-\frac{(P_{vv}-P_{ww}) (Tr P)}{2 \sqrt{2} a^7}
		\bigg)
			\nonumber
		\\
		&
	+\delta q_2 
	\bigg(
	\frac{k^2}{3a^3}
	-\frac{P_{kk}^2}{a^7}
	-\frac{P_{kv}^2}{a^7}
	-\frac{P_{kw}^2}{a^7}
	-\frac{ (Tr P)^2}{6 a^7}
	+\frac{ P_{kk} (Tr P)}{2a^7}
	+\frac{(Tr P^2)}{3 a^7}
	\bigg)
	+\frac{3 a V_{,\phi} \delta\phi}{2}
	\bigg]
		\nonumber
	\\
	&
	+\delta q_3 \bigg[
	\frac{2 P_{kk} P_{vw}\delta q_4}{a^7}
	+\frac{2 P_{kw} P_{vv}\delta q_5}{a^7}
	+\bigg(\frac{P_{kv} [(TrP)-P_{kk}]}{a^7}
	-\frac{2 P_{kw}	P_{vw}}{a^7}
	\bigg) \delta q_6
		\nonumber
	\\
	&
	-\frac{\sqrt{2}	P_{kv} \delta \pi_1}{3 a^3}
	-\frac{\sqrt{2} P_{kv} \delta\pi_2}{a^3}
	+\frac{P_{vw} \delta \pi_4}{a^3}
	+\frac{2 P_{kw} \delta \pi_5}{a^3}
	+\frac{2 P_{kv} \delta \pi_6}{a^3}
	+\delta \pi_3 \left(\frac{2(Tr P)}{3 a^3}
	-\frac{P_{ww}}{a^3}
	\right)
	\bigg]
		\nonumber
	\\
	&
	+\delta q_4 \bigg[
	\frac{2 P_{kv}P_{ww} \delta q_5}{a^7}
	+\bigg(
	\frac{2P_{kv} P_{vw}}{a^7}
	-\frac{P_{kw} [(TrP)-P_{kk}]}{a^7}
	\bigg)
	\delta q_6
	-\frac{\sqrt{2} P_{kw} \delta \pi_1}{3 a^3}
	-\frac{\sqrt{2}	P_{kw} \delta \pi_2}{a^3}
	+\frac{P_{vw} \delta \pi_3}{a^3}
	\nonumber
	\\
	&
	+\frac{2P_{kv} \delta \pi_5}{a^3}
	-\frac{2 P_{kw} \delta \pi_6}{a^3}
	+\delta\pi_4 \bigg(
	\frac{2 (Tr P)}{3 a^3}
	-\frac{P_{vv}}{a^3}
	\bigg)\bigg]
		\nonumber
	\\
	&
	+\delta q_5
	\bigg[
			\frac{(P_{vv}-P_{ww}) P_{vw}}{a^7}
			 \delta q_6
		-\frac{\sqrt{2} P_{vw} \delta \pi_1}{3a^3}
  		-\frac{\sqrt{2} P_{vw} \delta \pi_2}{a^3}
  		+\delta \pi_5
	\bigg(
	\frac{2 (Tr P)}{3 a^3}
	-\frac{P_{kk}}{a^3}\bigg)\bigg]
	\nonumber
	\\
	&
	+\delta q_6
	\bigg[
		-\frac{P_{vv}-P_{ww}}{3 \sqrt{2} a^3}
 \delta \pi_1
		-\frac{P_{vv}-P_{ww}}{\sqrt{2}a^3}
		 \delta \pi_2
		+\delta \pi_6	\bigg(\frac{2 (Tr P)}{3 a^3}
		-\frac{P_{kk}}{a^3}
		\bigg)\bigg]
			\nonumber
		\\
		&
	+\delta q_2
	\bigg[
	\frac{P_{kk} \delta \pi_2}{a^3}
	+\frac{4 \sqrt{2} P_{kv} \delta \pi_3}{3 a^3}
	+\frac{4 \sqrt{2} P_{kw} \delta \pi_4}{3 a^3}
	-\frac{2 \sqrt{2}P_{vw} \delta \pi_5}{3 a^3}
	-\frac{\sqrt{2} (P_{vv}-P_{ww})}{3	a^3}
     \delta \pi_6
	+\delta \pi_1	\left(\frac{(Tr P)}{9 a^3}
	-\frac{P_{kk}}{3 a^3}\right)
	\nonumber
	\\
	&
	+\delta q_3\bigg(\frac{\sqrt{2} P_{kk} P_{kv}}{3 a^7}
	-\frac{2 \sqrt{2} P_{vv}P_{kv}}{3 a^7}
	+\frac{\sqrt{2} (Tr P) P_{kv}}{3 a^7}
	-\frac{2	\sqrt{2} P_{kw} P_{vw}}{3 a^7}\bigg)
	\nonumber
	\\
	&
	+\delta q_4 \bigg(\frac{\sqrt{2}
		P_{kk} P_{kw}}{3 a^7}-\frac{2 \sqrt{2} P_{ww} P_{kw}}{3
		a^7}+\frac{\sqrt{2} (Tr P) P_{kw}}{3 a^7}-\frac{2 \sqrt{2}
		P_{kv} P_{vw}}{3 a^7}\bigg)
	\nonumber
	\\
	&
	+\delta q_5 
	\bigg(
	\frac{4 \sqrt{2} P_{kv}	P_{kw}}{3 a^7}
	-\frac{\sqrt{2} P_{kk} P_{vw}}{3a^7}
	-\frac{\sqrt{2} P_{vw} (Tr P)}{3 a^7}
	\bigg)
	\nonumber
	\\
	&
	+\delta q_6
	\bigg(
	\frac{2 \sqrt{2} P_{kv}^2}{3 a^7}
	-\frac{2 \sqrt{2} P_{kw}^2}{3a^7}
	-\frac{\sqrt{2} (P_{vv}^2-P_{ww}^2)}{3 a^7}
		\nonumber
	\\
	&
	-\frac{P_{kk} (P_{vv}-P_{ww})}{\sqrt{2} a^7}
	+\frac{(P_{vv}-P_{ww}) (Tr P)}{3 \sqrt{2}	a^7}
	\bigg)\bigg]
 .
\end{align}
\par
}

\section{Geometric quantities}\label{geoquant}

The canonical perturbation variables can be used to express geometric quantities. The most useful ones are the Ricci scalar,
\begin{align}
	\delta({^3}R)&
	=
	2a^{-4}
	k^2
	\left(
	\delta q_1
	-\frac{1}{3} \delta q_2
	\right),
\end{align}
the energy density of the scalar field,
\begin{align}
	\delta \rho=\frac{\mathcal{H}_{m,0}}{\sqrt{q}}\bigg|^{(1)}=- \frac{5 p_\phi^2}{4 a^8} \delta q_1+V_{,\phi} \delta \phi+\frac{p_\phi}{a^6} \delta \pi_\phi,
\end{align}
and the two scalar modes of the shear,
\begin{align}\begin{split}
	\delta\sigma_1&
	=	
	\frac{2}{9}a^{-2} K_2\delta q_2
	+\frac{1}{3}a^{-2} K_3\delta q_3
	+\frac{1}{3}a^{-2} K_4\delta q_4
	+\frac{1}{3}a^{-2} K_5\delta q_5
	+\frac{1}{3}a^{-2} K_6\delta q_6,\\
	\delta\sigma_{2}&
	=
	\frac{3}{2}a \delta \pi_2
	+\frac{3}{4} a^{-3}\delta q_1 
	\left(
	P_{kk} 
	-
	\frac{1}{3}(Tr P)
	\right)
	+ a^{-3}
	P_{kk} \delta q_2
	+\frac{1}{\sqrt{2}}a^{-3} P_{kv} \delta q_3
	\\&\qquad
	+\frac{1}{\sqrt{2}}a^{-3} P_{kw} \delta q_4
	+\sqrt{2}a^{-3} P_{vw} \delta q_5
	-\frac{1}{\sqrt{2}}a^{-3}( P_{vv}-P_{ww}) \delta q_6,
\end{split}
\end{align}
where $K_n=K_{ab}A^{ab}_n=a^{-3}\left(\bar{\pi}_{ab}A^{ab}_n-\frac{1}{2}\bar{q}_{ab}A^{ab}_n (Tr P) \right)$ are the components of the zeroth-order extrinsic curvature. 

\section{Physical Hamiltonian}\label{Hfin}

The coefficients in the physical Hamiltonian \eqref{hfin} read:

{\footnotesize \begin{align}\label{HfinCoeff}
	\tilde{U}_\phi&=
	\frac{(Tr P) p_\phi^2}{4 P_{kk} a^3}
	-\frac{3 p_\phi^2}{8a^3}
	-\frac{p_\phi V_{,\phi} a^3}{2 P_{kk}}
	+\frac{V_{,\phi\phi} a^3}{2};
	\nonumber
	\\
	\tilde{U}_5
	&=
	\frac{(Tr P)^2}{8a^7}
	+\frac{(Tr P^2)}{4 a^7}
	-\frac{( Tr P) P_{kk}}{2 a^7}
	+\frac{P_{kk}^2}{4a^7}
	+\frac{P_{kv}^2}{a^7}
	+\frac{P_{kw}^2}{a^7}
	-\frac{(P_{vv}-P_{ww})^2}{4a^7}
	\nonumber
	\\&
	\nonumber
	\quad
	+\frac{2P_{kv}P_{kw}P_{vw}}{a^7P_{kk}}
	-\frac{( Tr P) P_{vw}^2}{2 a^7P_{kk}}
	+\frac{P_{vw}^2}{4 a^7}
	+\frac{p_\phi^2}{8 a^7}
	-\frac{V}{4 a};
	\\
	\nonumber
	\tilde{U}_6
	&=
	\frac{(Tr P^2)}{4 a^7}
	+\frac{(Tr P)^2}{8a^7}
	-\frac{ (Tr P)P_{kk}}{2 a^7}
	+\frac{P_{kk}^2}{4 a^7}
	+\frac{P_{kv}^2}{a^7}
	+\frac{P_{kw}^2}{a^7}
	-\frac{P_{vw}^2}{a^7}
	\\&
	\nonumber
	\quad
	+\frac{(P_{kv}^2-P_{kw}^2)(P_{vv}-P_{ww})}{2a^7P_{kk}}
	-\frac{(Tr P)(P_{vv}-P_{ww})^2}{8a^7P_{kk}}
	+\frac{(P_{vv}-P_{ww})^2}{16a^7}
	+\frac{p_\phi^2}{8 a^7}
	-\frac{V}{4 a};
	\\
	\nonumber
	\tilde{C_1}&=
	-\frac{(Tr P)P_{vw}(P_{vv}-P_{ww})}{2a^7 P_{kk}}+\frac{5P_{vw}(P_{vv}-P_{ww})}{4a^7}+\frac{P_{vw}(P_{kv}^2-P_{kw}^2)}{a^7 P_{kk}}
	\\&
	\nonumber
	\quad
	+\frac{P_{kv} P_{kw}(P_{vv}-P_{ww})}{a^7P_{kk}};
	\\
	\tilde{C}_2&=
	\frac{\sqrt{2} P_{kv} P_{kw} p_\phi}{a^5	P_{kk}}
	-\frac{ P_{vw} p_\phi}{2 \sqrt{2} a^5}
	-\frac{a P_{vw} V_{,\phi}}{\sqrt{2}	P_{kk}};
	\\
	\nonumber
	\tilde{C}_3&=
	\frac{ (P_{kv}^2-P_{kw}^2)p_\phi}{\sqrt{2} a^5 P_{kk}}
	-\frac{(P_{vv}-P_{ww})p_\phi}{4 \sqrt{2} a^5}
	-\frac{a (P_{vv}-P_{ww})V_{,\phi}}{2 \sqrt{2} P_{kk}};
	\\
	\nonumber
	C_{\phi\phi}&=-\frac{p_{\phi}^2}{2a^3 P_{kk}} ;
	\\
	\nonumber
	C_{55}&=\frac{1}{a^3}\bigg[\frac{2}{3}(Tr P)-P_{kk}-\frac{2P_{vw}^2}{P_{kk}}\bigg];
	\\
	\nonumber
	C_{66}&=\frac{1}{a^3}\bigg[\frac{2}{3}(TrP)-P_{kk}-\frac{(P_{vv}-P_{ww})^2}{2P_{kk}}\bigg];
	\\
	\nonumber
	C_{5\phi}&=-\frac{P_{vw} p_{\phi}}{a^3 P_{kk}};
	\\
	\nonumber
	C_{6\phi}&=-\frac{(P_{vv}-P_{ww})p_\phi}{2 a^3 P_{kk}};
	\\
	\nonumber
	C_{56}&=-\frac{P_{vw}(P_{vv}-P_{ww})}{a^3P_{kk}}.
\end{align}}

\section{Dynamics of the operator $P$}\label{Appbackeom}
Making use of Eqs (\ref{backeom}) and (\ref{fermilaw}) we find the dynamics of the components of the operator $P$ in the Fermi-Walker basis:
\begin{align}\begin{aligned}
\frac{\ud}{\ud t}P_{kk}=a^{-3}\left(-a^6V-2P_{kv}^2-2P_{kw}^2\right),
&~~\frac{\ud}{\ud t}P_{vv}=a^{-3}\left(-a^6V+2P_{kv}^2\right),
\\
\frac{\ud}{\ud t}P_{ww}=a^{-3}\left(-a^6V+2P_{kw}^2\right),
&~~\frac{\ud}{\ud t}P_{vw}=2a^{-3}P_{kw}P_{kv}
,
\\
\frac{\ud}{\ud t}P_{kv}=a^{-3}\left(P_{kk} P_{kv}-P_{kv}P_{vv}-P_{kw}P_{vw}\right),
&~~
\frac{\ud}{\ud t}P_{kw}=a^{-3}\left(P_{kk} P_{kw}-P_{kw}P_{ww}-P_{kv}P_{vw}\right).
\end{aligned}
\end{align}

\section{Final Hamiltonian in the M-S variables}\label{MSAppgen}

Below the coefficients in the final Hamiltonians (\ref{HBI}), (\ref{BIn}) and (\ref{HBIgen}) are given. In case of a single scalar field there is a unique field label $I=J$ which is omitted in the Hamiltonian.

\begin{align}
&U_{\phi^I}=\frac{p_{\phi^I}^2(P_{kv}^2 + P_{kw}^2 +P_{kv}P_{kw}+P_{vw}(P_{kv}+P_{kw})-P_{vw}^2+a^6V)}{a^4 P_{kk}^2}-\frac{2a^6p_{\phi^I}V_{,\phi^I}}{a^4 P_{kk}}
\nonumber
\\
&+\frac{(Tr P)^2-18a^6(V-2V_{,\phi^I\phi^I})}{36 a^4};\nonumber\\
&U_5=\frac{(Tr P)^2+72 \left(P_{kv}^2+P_{kv} P_{kw}+P_{kw}^2+ P_{vw} (P_{kv}+P_{kw})-P_{vw}^2\right)+18 \sum_I p_{\phi^I}^2+18 a^6 V}{36a^4}\nonumber\\
&+\frac{P_{vw}^2 (4 P_{kv}^2 + 4 P_{kw}^2 - 4 P_{vw}^2 - (P_{vv} - P_{ww})^2 -P_{kk}^2- \sum_Ip_{\phi^I}^2 + 
   2 a^6 V)}{a^4P_{kk}^2}-2\frac{\frac{(P_{vv}-P_{ww})^2}{4}}{a^4}\nonumber\\
   &+\frac{54 P_{kk}^4 - 36 P_{kk}^3 (Tr P)}{36a^4P_{kk}^2}+\frac{2 P_{vw} (8 P_{kv} P_{kw} + P_{vw} (Tr P))}{a^4P_{kk}};\nonumber\\
 &U_6=\frac{(Tr P)^2 + 72 (P_{kv}^2 + P_{kv} P_{kw} +P_{kw}^2+ P_{vw}( P_{kv}+P_{kw} )-P_{vw}^2) + 
 18 \sum_Ip_{\phi^I}^2 + 18 a^6 V}{36a^4}\nonumber\\
 &+\frac{\frac{(P_{vv} - P_{ww})^2}{4} (4 P_{kv}^2 + 4 P_{kw}^2 - 4 P_{vw}^2- (P_{vv} - P_{ww})^2-P_{kk}^2 -
    \sum_Ip_{\phi^I}^2 + 2 a^6 V)}{a^4P_{kk}^2}-2\frac{P_{vw}^2}{a^4}\nonumber\\
    &+\frac{54 P_{kk}^4 -36 P_{kk}^3(Tr P) }{36a^4P_{kk}^2}+\frac{ 2\frac{(P_{vv} - P_{ww})}{2} (4 P_{kv}^2 - 4 P_{kw}^2 +\frac{(P_{vv} - P_{ww})}{2}(Tr P))}{a^4P_{kk}};\nonumber\\
&
C_1=\frac{ 
  P_{vw} (P_{vv} - P_{ww}) (4 P_{kv}^2 + 4 P_{kw}^2 - 4 P_{vw}^2 - (P_{vv} - P_{ww})^2 - 
     \sum_Ip_{\phi^I}^2 + 2 a^6 V)}{a^4P_{kk}^2}+
 \nonumber
 \\
 &
  \frac{8(P_{kv}^2 - P_{kw}^2)P_{vw}+ 
     8 P_{kv} P_{kw} (P_{vv} - P_{ww}) + 2(Tr P) P_{vw} (P_{vv} - P_{ww})}{a^4P_{kk}}+\frac{P_{vw} (P_{vv} - P_{ww}) }{a^4};\nonumber\\
 &
 C_{2I}=\frac{P_{vw} p_{\phi^I} (4 P_{kv}^2 + 4 P_{kw}^2 - 4 P_{vw}^2 - (P_{vv} - P_{ww})^2 - 
     \sum_Ip_{\phi^I}^2 + 2 a^6 V) }{a^4P_{kk}^2}-\frac{3 P_{vw} p_{\phi^I}}{a^4}
\nonumber \\
     &
     +\frac{8 P_{kv} P_{kw}p_{\phi^I} + 2(Tr P) P_{vw} p_{\phi^I} -4 a^6 P_{vw} V_{,\phi^I}}{a^4P_{kk}};\nonumber\\
     &
C_{3I}=\frac{\frac{(P_{vv}-P_{ww})}{2} p_{\phi^I} (4 P_{kv}^2 + 4 P_{kw}^2 - 4 P_{vw}^2 - (P_{vv} - P_{ww})^2 - 
     \sum_Ip_{\phi^I}^2 + 2 a^6 V) }{a^4P_{kk}^2}-\frac{3 \frac{(P_{vv} - P_{ww})}{2} p_{\phi^I}}{a^4}
\nonumber \\
     &
     +\frac{ p_{\phi^I}(4 P_{kv}^2- 4P_{kw}^2 + (Tr P) (P_{vv}-P_{ww})) -2a^6 (P_{vv}-P_{ww}) V_{,\phi^I}}{a^4P_{kk}};\nonumber\\
     &
 C_{IJ} (I\neq J)=\frac{ \left(4 a^6 V_{,\phi^I\phi^J}-3 p_{\phi^I} p_{\phi^J}\right)}{2 a^4}-\frac{2 a^6 (p_{\phi^J} V_{\phi^I}+p_{\phi^I} V_{\phi^J})}{ a^4 P_{kk}}-\frac{(Tr P) p_{\phi^I} p_{\phi^J}}{a^4 P_{kk}}
 \nonumber \\
     &
     -\frac{p_{\phi^J} p_{\phi^I} \left(-2 a^6 V-4 P_{kv}^2-4 P_{kw}^2+(P_{vv}-P_{ww})^2+4 P_{vw}^2+\sum_Ip_{\phi^I}^2\right)}{2 a^4 P_{kk}^2}.
\end{align}

\section{Dirac observables}\label{Dext}
This is a complete set of solutions to Eq. (\ref{defdir}). In case of a single scalar field the label $I$ is unique and can be omitted.
\begin{align}
	&{\delta D_1}
	=
	{\delta q _5}
	+
	2\sqrt{2}
	\frac{P_{vw}}{P_{kk}}
	{\delta q_1}
	-
	\frac{2\sqrt{2}}{3}
	\frac{P_{vw}}{P_{kk}}
	{\delta q _2}
	;
	\\&
	{\delta D_2}
	=	
	{\delta q _6}
	+
	\frac{\sqrt{2}}{P_{kk}}
	\big(
	P_{vv}
	-
	P_{ww}
	\big)
	{	\delta q_1}
	-
	\frac{\sqrt{2}}{3}
	\frac{1}{P_{kk}}
	\big(
	P_{vv}
	-
	P_{ww}
	\big)
	{	\delta q _2}
	;
	\\&
	{\delta D_3}
	=	
	{	\delta \pi_1}
	+
	\bigg\{
	\frac{1}{P_{kk}}
	\bigg[
	2
	k^2
	-
	\frac{1}{2}
	a^{-4}
	(Tr P^2)
	+
	\frac{3}{4}
	a^{-4}
	(Tr P)^2
+4	a^{-4}(P_{kv}^2+ P_{kw}^2)
	+\frac{3}{4}
	a^{-4}
	\sum_Ip^2_{\phi^I}
	-
	\frac{3}{2}
	a^2
	V(\phi)	
	\bigg]
		\nonumber\\
	&
	\qquad\quad
	+3
	a^{-4}
	P_{kk} 
	-\frac{5}{2}
	a^{-4}
	(Tr P) 
	\bigg\}
	{	\delta q_1}
	\nonumber\\
	&
	\qquad\quad
	+
	\frac{1}{3P_{kk}}
	\bigg[
	-
	2
	k^2
	+
	\frac{1}{2}
	a^{-4}
	(Tr P^2)
	-
	\frac{3}{4}
	a^{-4}
	(Tr P )^2
		+
	a^{-4}
	(Tr P)
	P_{kk}
	-\frac{3}{4}
	a^{-4}
	\sum_Ip^2_{\phi^I}
	+
	\frac{3}{2}
	a^2
	V(\phi)
	\nonumber\\
	&
	\qquad\quad
	-
	4
	a^{-4}
	(
		P_{kv}^2
		+
	P_{kw}^2)
	\bigg]
	{	\delta q _2}
	+
	{\sqrt{2}}
	a^{-4}
	P_{kv}
	{	\delta q _3}
	+
	{\sqrt{2}}
	a^{-4}
	P_{kw}
	{\delta q _4};
	\\
	&
	{\delta D_4}
	=
	{	\delta \pi_2}
	+
	\bigg\{
	\frac{1}{P_{kk}}
	\bigg[
	-\frac{2}{3}
	k^2
	+
	\frac{2}{3}
	a^{-4}
	(Tr P^2)
	-
	\frac{1}{2}
	a^{-4}
	(Tr P)^2
	-
	2a^{-4}
	P_k^2
	-
	\frac{4}{3}
	a^{-4}
	(	P_{kv}^2+P_{kw}^2)
		\bigg]
	\nonumber\\
	&
	\qquad\quad
	+
	a^{-4}
	\frac{1}{2} 
	P_{kk}
	+
	a^{-4} \frac{4}{3} 
	(Tr P )
	\bigg\}
	{	\delta q_1}
	\nonumber\\
	&
	\qquad\quad
	+
	\frac{1}{3P_{kk}}
	\bigg[
	\frac{2}{3}
	k^2
	+
	2a^{-4}
	P_{kk}^2
	-
	\frac{2}{3}
	a^{-4}
	(Tr P^2)
	-\frac{5}{6}
	a^{-4}
	(Tr P)
	P_{kk}
	\nonumber\\
&
\qquad\quad
	+
	\frac{1}{2}
	a^{-4}
	(Tr P)^2
	+	\frac{10}{3}
	a^{-4}
(	P_{kw}^2
	+
	P_{kv}^2)
	\bigg]
	{	\delta q _2}
	-\frac{\sqrt{2}}{3}a^{-4}
	P_{kv}	
	{	\delta q _3}
	-\frac{\sqrt{2}}{3}a^{-4}
	P_{kw}
	{	\delta q _4};
	\\
	&
	{\delta D_5}=	
	{\delta \pi_3}
	-
	\frac{1}{P_{kk}}
	\bigg(
	2{\sqrt{2}
		a^{-4}
		(P_{kv}P_{vv}+P_{kw}P_{vw})
	}
	-
	\sqrt{2}
	a^{-4}
	(Tr P)
	P_{kv}
	\bigg)
	{	\delta q_1}
	\nonumber\\
	&
	\qquad\quad
	+
	\frac{1}{3P_{kk}}
	\bigg(
	2{\sqrt{2}}
	a^{-4}
		(P_{kv}P_{vv}+P_{kw}P_{vw})
	-
	\sqrt{2}
	a^{-4}
	(Tr P)
	P_{kv}
	\bigg)
	{	\delta q _2}
	+
	a^{-4}
	P_{kk} 
	{	\delta q _3};
	\\
	&
	{\delta D_6}=
	{\delta \pi_4}	
	-
	\frac{1}{P_{kk}}
	\bigg(
	2\sqrt{2}
	a^{-4}
		(P_{kv}P_{vw}+P_{kw}P_{ww})
	-
	\sqrt{2}
	a^{-4}
	(Tr P)
	P_{kw}
	\bigg)
	{	\delta q_1}
	\nonumber\\
	&
	\qquad\quad
	+
	\frac{1}{3P_{kk}}
	\bigg(
	2
	\sqrt{2}
	a^{-4}
		(P_{kv}P_{vw}+P_{kw}P_{ww})
	-
	\sqrt{2}
	a^{-4}
	(Tr P)
	P_{kw}
	\bigg)
	{	\delta q _2}
	+
	a^{-4}
	P_{kk} 
	{	\delta q _4};
	\\
	&
	{\delta D_7}
	=
	{\delta \pi_5}
	+
	\bigg[
	\frac{1}{P_{kk}}
	\bigg(
		2\sqrt{2}a^{-4}	
	P_{kw}P_{kv}	
	-\frac{1}{\sqrt{2}} 
	a^{-4}(Tr P)
	P_{vw} 	
	\bigg)
	+\frac{1}{\sqrt{2}} 
	a^{-4}
	P_{vw} 
	\bigg]
	{	\delta q_1}
			\nonumber\\
	&
	\qquad\quad
	+
	\frac{1}{3P_{kk}}
	\bigg(
	\frac{1}{\sqrt{2}} 
	a^{-4}(Tr P)
	P_{vw}
	-2
	\sqrt{2}
		a^{-4}
	(P_{kk}P_{vw}
	-
	P_{kw}P_{kv})
	\bigg)
	{	\delta q _2}
			\nonumber\\
	&
	\qquad\quad
	+
	a^{-4}
	P_{kw}
	{	\delta q _3}
	+
	a^{-4}
	P_{kv}
	{	\delta q _4};
	\\
	&
	{\delta D_8}
	=	
	{\delta \pi_6}
	+
	\bigg\{
	\frac{1}{P_{kk}}
	\bigg[
	-\sqrt{2}
	a^{-4}
	(P_{vv}^2-P_{ww}^2)
		+
	\frac{3}{2\sqrt{2}}
	a^{-4}
	(Tr P)
	(P_{vv}-P_{ww})
	+
	\sqrt{2}	a^{-4}	
	(P_{kv}^2-P_{kw}^2)
	\bigg]
			\nonumber\\
	&
	\qquad\quad
	-\frac{3}{2\sqrt{2}} a^{-4}
(	P_{vv}
-
	P_{ww})
	\bigg\}
	{	\delta q_1}
	+
	\frac{1}{3P_{kk}}
	\bigg[
	\sqrt{2}
	a^{-4}
	(P_{vv}^2-P_{ww}^2)
	-
	\frac{3}{2\sqrt{2}}
	a^{-4}
	(Tr P)
	(P_{vv}-P_{ww})
		\nonumber\\
	&
	\qquad\quad
	-
	\sqrt{2}	a^{-4}	
	(P_{kv}^2-P_{kw}^2)	
	\bigg]
	{	\delta q _2}
	+
	a^{-4}
	P_{kv}	
	{	\delta q _3}
	-
	a^{-4}
	P_{kw}
	{	\delta q _4};
		\\
	&
	{\delta D_{9I}}
	=
	{\delta \phi^I}
	+
	\frac{1}{P_{kk}}
	a^{-2}
	p_{\phi^I}
	{	\delta q_1}
	-
	\frac{1}{3P_{kk}}
	a^{-2}
	p_{\phi^I}
	{	\delta q _2};
	\\
	&
	{\delta D_{10I}}
	=	
	{\delta \pi_{\phi^I}}
	+
	\frac{1}{a^2}\bigg[
	\frac{1}{ P_{kk}}
	\bigg(
	\frac{1}{2}(Tr P)
	p_{\phi^I}
	-a^6
	V_{,\phi^I}
	\bigg)	
	-\frac{3}{2}
	p_{\phi^I}	
	\bigg]
	{	\delta q_1}
	+
	\frac{1}{3a^2P_{kk}}
	\bigg(a^6
	V_{,\phi^I}
	-\frac{1}{2}
	(Tr P)
	p_{\phi^I}\bigg)	
	{	\delta q _2};
\end{align}

\section{Geometric expressions for the Dirac observables}\label{geodirac}

The independent Dirac observables (\ref{inddir}) may be also given in terms of geometric quantities,
\begin{align}
	&\delta Q_1=
	\frac{1}{\sqrt{2} a}\delta q_5
	-
	\frac{a^3K_5}{2\sqrt{2}k^2\left(K_1-\frac{1}{3}K_2\right)}
	\delta R,
	\nonumber\\
	& \delta Q_2=
	\frac{1}{\sqrt{2} a}\delta q_6
	-\frac{a^3K_6}{2\sqrt{2}k^2\left(K_1-\frac{1}{3}K_2\right)}
	\delta R,
	\nonumber\\
	& \delta Q_3=a\delta\phi
	-\frac{a^2p_{\phi}}{4k^2\left(K_1-\frac{1}{3}K_2\right)}\delta R,
	\nonumber\\
	&\delta P_1=\sqrt{2} \delta K_5
	+\frac{K_5 K_6}{4\sqrt{2} a^2\left(K_1-\frac{1}{3}K_2\right)}\delta q_6
	+\frac{K_1\left(K_1-\frac{1}{3}K_2\right)
		-2\left(K_1-\frac{1}{3}K_2\right)^2 
		+\frac{1}{2}K_5^2}
	{ \sqrt{2} a^2\left(K_1-\frac{1}{3}K_2\right)}\delta q_5,
	\nonumber\\
	&+\mathcal{G}\left(\frac{aK_5}{\sqrt{2}},\frac{a^2K_3K_4}{2} \right)\delta R
	+\frac{K_5 p_{\phi}}{2a\sqrt{2} \left(K_1-\frac{1}{3}K_2\right)}\delta\phi
	\\
	& \delta P_2=\sqrt{2} \delta K_6
	+\frac{K_5 K_6}{4\sqrt{2} a^2\left(K_1-\frac{1}{3}K_2\right)}\delta q_5
	+\frac{ 4K_1\left(K_1-\frac{1}{3}K_2\right)
		-8\left(K_1-\frac{1}{3}K_2\right) ^2
		+\frac{1}{2}K_6^2}
	{4 \sqrt{2} a^2 \left(K_1-\frac{1}{3}K_2\right)}\delta q_6
	\nonumber\\
	&+\mathcal{G}\left(
	\frac{aK_6}{2\sqrt{2}},
	\frac{a^2(K_3^2-K_4^2)}{4}\right)
	\delta R+\frac{p_{\phi} K_6}{4\sqrt{2} a \left(K_1-\frac{1}{3}K_2\right)}\delta\phi
	\nonumber\\
	& \delta P_3=
	\frac{a^5}{p_{\phi}}\delta\rho
	+\frac{K_5 p_{\phi}}{2 a^3\left(K_1-\frac{1}{3}K_2\right)}\delta q_5
	+\frac{K_6 p_{\phi}}{8 a^3 \left(K_1-\frac{1}{3}K_2\right)}\delta q_6
	\nonumber\\&
	+\frac{12a^2K_1\left(K_1-\frac{1}{3}K_2\right)p_{\phi}
		+3 p_{\phi}^3
		-12a^7\left(K_1-\frac{1}{3}K_2\right)V_{,\phi}}
	{12a^2 p_{\phi}\left(K_1-\frac{1}{3}K_2\right)}\delta\phi
	\nonumber\\
	&+\frac{8a^3p_{\phi} K_1\left(K_1-\frac{1}{3}K_2\right)-
		a p_{\phi} \left(\frac{a^2}{2}K_6^2
		+2a^2K_5^2+p_{\phi}^2\right)
		+4 a^8 \left(K_1-\frac{1}{3}K_2\right) V_{,\phi}}{16 k^2a^2 \left(K_1-\frac{1}{3}K_2\right)^2}\delta R,
	\nonumber
\end{align}
where $\mathcal{G}(X,Y)=a\frac{-24a^2\left(K_1-\frac{1}{3}K_2\right)^2X -4a\left(K_1-\frac{1}{3}K_2\right) (6 Y -6a K_1 X) - 
	3 X (2 a^2K_5^2+ \frac{a^2}{2}K_6^2 + p_{\phi}^2)}{6k^2 P_{kk}^2}$. In the isotropic limit we obtain:

\begin{equation}
	\begin{gathered}
	\delta Q_1\rightarrow\frac{1}{\sqrt{2} a}\delta q_5,
	\quad
	 \delta Q_2\rightarrow\frac{1}{\sqrt{2} a}\delta q_6,
	\quad
	\delta Q_3\rightarrow 
	a\delta\phi-\frac{a^2p_{\phi}}{4k^2K_1}\delta R,
\\
	\delta P_1\rightarrow\sqrt{2} \delta K_5 -
	\frac{K_1}{ \sqrt{2} a^2}\delta q_5,
	\quad
	 \delta P_2\rightarrow\sqrt{2} \delta K_6
	-\frac{K_1}{ \sqrt{2} a^2}\delta q_6,
\\
	 \delta P_3\rightarrow\frac{a^5}{p_{\phi}}\delta\rho
	+\frac{36a^2K_1^2p_{\phi}+9 p_{\phi}^3-36a^7K_1V_{,\phi}}
	{36 a^2 p_{\phi}K_1}\delta\phi
	+\frac{72a^3p_{\phi} K_1^2-9a p_{\phi}^3+36 a^8 K_1V_{,\phi}}{144 k^2 a^2K_1^2}\delta R.
		\end{gathered}
\end{equation}

\section{Comparison of metric decompositions}\label{Uzantensor}

The perturbed metric decomposition used in \cite{Uzan} is given by

\begin{align}\label{hij}
	h_{ij}=2C\left(\gamma_{ij}+\frac{\sigma_{ij}}{\mathcal{H}}\right)+2\partial_i\partial_j E+2\partial_{(i}E_{j)}+2E_{ij},
\end{align}
whereas our decomposition reads
\begin{align}\label{deltagamma}
	h_{ij}=a^{-2}\delta q_{ij}=a^{-2}\delta q_n A^n_{ij}.
\end{align}
Expanding \eqref{hij} in the $A$ basis,
\begin{align}\label{hijcoeff}
	h_{ij}&
	=
	A^1_{ij}\left(2C-\frac{2}{3}E\right)
	+
	A^2_{ij}\left(2C\frac{\sigma_2}{\mathcal{H}}-2E\right)
	+
	A^3_{ij}\left(2C\frac{\sigma_3}{\mathcal{H}}+i\sqrt{2}E_v\right)
	+
	A^4_{ij}\left(2C\frac{\sigma_4}{\mathcal{H}}+i\sqrt{2}E_w\right)
	\nonumber\\&\qquad\qquad
	+
	A^5_{ij}\left(2C\frac{\sigma_5}{\mathcal{H}}+2E_5\right)
	+
	A^6_{ij}\left(2C\frac{\sigma_6}{\mathcal{H}}+2E_6\right),
\end{align}
and comparing with \eqref{deltagamma} we obtain the relation between the tensor modes:
\begin{align}
	E_5&
	=\frac{a^{-2}}{2} \delta q_5
	+
	a^{-2}\sqrt{2}
	\frac{P_{vw}}{P_{kk}}
	\left(\delta q_1 -\frac{1}{3}\delta q_2\right)
,\quad
	E_6
	=\frac{a^{-2}}{2}\delta q_6
	+\frac{a^{-2}(P_{vv}-P_{ww})}{\sqrt{2}P_{kk}}\left(\delta q_1 -\frac{1}{3}\delta q_2\right),
\end{align}
which correspond to the gauge-invariant quantities $\delta Q_1$ and $\delta Q_2$ given in Eq. \eqref{inddir}. For the vector and scalar modes we find the following relations:
\begin{align}\begin{split}
	E_v&=-i\frac{2P_{kv}}{P_{kk}}a^{-2}
	\left(\delta q_1 -\frac{1}{3}\delta q_2\right)
	-i \frac{a^{-2}}{\sqrt{2}}\delta q_3
,\quad
	E_w
	=-i\frac{2	P_{kw}}{P_{kk}}a^{-2}
	\left(\delta q_1 -\frac{1}{3}\delta q_2\right)
	-i\frac{1}{\sqrt{2}}a^{-2}\delta q_4,\\
	E
	&=
	\frac{a^{-2}(TrP)}{2P_{kk}}
	\left(
	\delta q_1
	-\frac{1}{3}\delta q_2
	\right)
	-\delta q_1a^{-2}\frac{3}{2}
,\quad
	C
	=
	\frac{a^{-2}(TrP)}{6P_{kk}}
	\left(\delta q_1 -\frac{1}{3}\delta q_2\right).\end{split}
\end{align}

\newpage

\bibliography{References}

\end{document}